\documentclass[twocolumn,pra,aps,superscriptaddress,showpacs]{revtex4-1}

 \usepackage{mathptmx}
\usepackage{dcolumn}
\usepackage{amsmath,amssymb,amsfonts}
\usepackage{dsfont}
\usepackage{bm}
\usepackage{color}
\usepackage{latexsym}
\usepackage{epstopdf}
\usepackage[english]{babel}
\usepackage{enumerate}

\usepackage[pdftex]{graphicx}

\usepackage{natbib}
\usepackage{epstopdf}
\usepackage{appendix}

\definecolor{mygrey}{gray}{0.35}
\definecolor{myblue}{rgb}{0.2,0.2,0.8}
\definecolor{myzard}{cmyk}{0,0,0.05,0}
\definecolor{mywhite}{rgb}{1,1,1}
\definecolor{mywhite}{rgb}{1,1,1}
\definecolor{myred}{rgb}{1,0.,0.3}
\usepackage[colorlinks=true,citecolor=myblue,linkcolor=myred]{hyperref}

\newcommand{\bra}[1]{\left\langle #1\right|}
\newcommand{\ket}[1]{\left| #1\right\rangle}
\newcommand{\braket}[2]{\langle #1|#2\rangle}

\newcommand\aaa{\mathbf{a}}
\newcommand\bb{\mathbf{b}}

\newcommand\hh{\mathbf{h}}
\newcommand\KKK{\mathbf{K}}

\newcommand\jj{\mathbf{j}}
\newcommand\nn{\mathbf{n}}
\newcommand\mm{\mathbf{m}}

\newcommand\kk{\mathbf{k}}
\newcommand\qq{\mathbf{q}}
\newcommand\pp{\mathbf{p}}

\newcommand\KK{K}

\newcommand\intt{\mathrm{int}}

\newcommand\BS{\mathrm{BS}}

\begin{document}

\title{Exotic Quantum Dynamics and Purely Long-Range Coherent Interactions in Dirac Cone-like Baths.}
 \author{A. Gonz\'{a}lez-Tudela}
 \email{alejandro.gonzalez-tudela@mpq.mpg.de}
 \affiliation{Max-Planck-Institut f\"{u}r Quantenoptik Hans-Kopfermann-Str. 1. 85748 Garching, Germany }
 \author{J. I. Cirac}
 \affiliation{Max-Planck-Institut f\"{u}r Quantenoptik Hans-Kopfermann-Str. 1. 85748 Garching, Germany }

\begin{abstract}
 In this work we study the quantum dynamics emerging when quantum emitters exchange excitations with a two-dimensional bosonic bath with a Dirac-Cone dispersion. We show that a single quantum emitter spectrally tuned in the vicinity of the Dirac point relaxes following a logarithmic law in time. Moreover, when several emitters are coupled to the bath at that frequency, long-range coherent interactions between them appear which decay inversely proportional to their distance without exponential attenuation. We analyze both the finite and infinite system situation using both perturbative and non-perturbative methods.
\end{abstract}

\maketitle

\section{Introduction}

Recently, the interest on the dynamics of quantum emitters (QEs) coupled to a structured baths has revived due to the prospects offered by new experimental platforms~\cite{vetsch10a,huck11a,hausmann12a,laucht12a,thompson13a,goban13a,beguin14a,lodahl15a,bermudez15a,sipahigi16a,corzo16a,sorensen16a,sipahigi16a,solano17a,devega08a,navarretebenlloch11a,houck12a,astafiev10a,hoi11a,vanloo13a,liu17a}. Of particular interest are the way in which the dynamics of the QE is modified by the bath~(see \cite{purcell46a,nakazato96a,lambropoulos00a,woldeyohannes03a,giraldi11a,breuer16a,devega17a} and references therein), the way the latter can mediate interactions or collective dissipation~\cite{ kien05a,dzsotjan10a,gonzaleztudela11a,chang12a,shahmoon13a,douglas15a,gonzaleztudela15c,shahmoon16a} or how the QEs can induce entanglement in the bath~\cite{gonzaleztudela15a,gonzaleztudela16a,douglas16a}.

For a single QE, unconventional relaxations dynamics have been predicted when the QE energies lie close to band-edges~\cite{john94a} or close to 2D Van-Hove singularities in the middle of the band~\cite{gonzaleztudela17a,gonzaleztudela17b}. The former is associated to the emergence of atom-photon bound states~\cite{bykov75a,john90a,kurizki90a}, which leads to fractional decay, whereas the latter is also accompanied by a highly directional emission into the bath~\cite{gonzaleztudela17a,gonzaleztudela17b,galve17a,mekis99a,langley96a}, and very slow relaxation dynamics scaling with $\sim 1/(t^2\log(t)^2)$.

For two (or more) QEs, the possibility of obtaining long-range coherent interactions between quantum emitters can be harnessed, e.g., to generate long-distance entanglement~\cite{shahmoon13a}, but also to simulate spin-models with long-range interactions. Non-conventional phenomena occur when the interactions decay with the distance, $r$, as $1/r^a$, with $a<D$ and $D$ being the spatial dimension on the bath. For instance, non-local transmission of correlations \cite{hauke13a,richerme14a,maghrebi16a}, violation of the area law~\cite{koffel12a}, fast equilibration~\cite{kastner11a} and quantum state transfer~\cite{vodola14a,eldredge16a}.

The coupling to the three-dimensional structureless photonic bath ($D=3$) induces coherent interactions decaying with the distance as a power law~\cite{lehmberg70a,lehmberg70b}. However, they are unavoidably accompanied by dissipative terms of the same order which compete with them. A way of avoiding such dissipative terms consists in using \emph{structured} reservoirs, such as photonic band-gap materials \cite{joannopoulos_book95a}, and exploit the emergence of atom-photon bound states~\cite{bykov75a,john90a,kurizki90a} to mediate the interactions~\cite{douglas15a,douglas16a}. 
The price to pay in this case for the cancellation of the dissipative terms, however, is that the interactions are ultimately exponentially attenuated by the length of the bound states. Even though this length can be large, it is of fundamental and practical interest to know whether there is a bath that can combine the best of both scenarios, that is, having no dissipative terms as photonic band-gap materials, while keeping unattenuated interactions like 3D structureless baths.  Inspired by pioneering works on graphene (see \cite{pereira06a,castroneto09a} and references therein), in this manuscript we provide a positive answer to this  question. In particular, we show how 
by coupling 
the QEs to a bosonic bath with a Dirac cone dispersion relation gives rise to such scenario.

In the context of graphene research~\cite{castroneto09a}, transport of energies around the Dirac cone in the presence of classical scatterers has been thoroughly analyzed. In that problem, a localized state emerging around the scatterers plays a crucial role. Its nature is very special given that its wavefunction decays very slowly with the distance, $r$, to the scatterer [$\sim1/r$]. This function is (marginally) not square-integrable in 2D, and thus it is labeled as \emph{quasi-bound state} (qBS)~\cite{pereira06a}. As we will show here, such state also has crucial consequences when instead of scatterers we have QEs, and instead of a fermionic bath, we replace it by a bosonic one.

In this manuscript, we solve exactly the quantum dynamics of initially excited QEs with their transition frequency exactly at the Dirac point.  We compare the results with those predicted by perturbative methods based on Born-Markov (or single-pole approximations)~\cite{lehmberg70a,lehmberg70b} that are normally used in quantum optics. For a single QE, Markov approximation predicts no decay because the density of states is zero, whereas the exact dynamics predicts a logarithmic relaxation to the ground state in the thermodynamic limit; that is,  when the number of modes, $N$, in the bath diverges. This very slow relaxation quenches at a certain time for finite systems, leading to a fractional decay of excitations. For two QEs coupled to different sublattices, Born-Markov approximation predicts purely long-range coherent interactions with $a=1$ and \emph{without} exponential attenuation in the thermodynamic limit, whereas the exact treatment predicts no interactions at all. We characterize the convergence to 
the thermodynamic limit showing 
that the exact dipole-dipole interactions decay with a $1/\log(N)
$ dependence. Thus, for practical purposes these 
interactions can still be harnessed in large systems as we show numerically. We provide both analytical and numerical understanding of the phenomena and the interpolation between 
finite and infinite systems, which allows us to evaluate to which extent these interactions can be observed.

The manuscript is structured as follows: in Section~\ref{sec:system} we describe the system under study by writing explicitly all the terms of the Hamiltonian and explain the bath properties. Then, in Section~\ref{sec:single}, we characterize the dynamics of a single QE in both finite and infinite systems. In Section~\ref{sec:two} we study the physics emerging when two QEs are coupled to the bath, generalizing the conclusions to many QEs. In Section~\ref{sec:experiment} we give a brief overview of several experimental limitations, and how they will affect to the observation of the phenomen. Finally, we conclude the paper in Section~\ref{sec:conclu} by summarizing the main results and pointing to future directions of work.

\section{System \label{sec:system}}

For the bath, we use a two-band model formed by two different lattices, that we denote as A/B lattices. The annihilation operators of the A/B lattice modes are described in terms of bosonic operators, $a_\nn$ and $b_\nn$, respectively.  We assume both lattices to be degenerate in energy $\omega_a=\omega_b$. The A/B lattices are connected through nearest-neighbor coupling $J$, such that the bath Hamiltonian reads as follows (using $\hbar=1$): $H_B=J\sum_{\nn}\sum_{\mm=0,\aaa_1,\aaa_2}\left(a^\dagger_{\nn+\mm} b_{\nn}+\mathrm{h.c.}\right)$. The positions within the A/B lattices are given by two integer numbers $n_{1,2}=1,\dots, N$, such that $\nn=(n_1,n_2)\equiv n_1 \aaa_1+n_2\aaa_2$, where $\aaa_{1,2}$ are the primitive lattice vectors, e.g., for hexagonal lattices $\aaa_{1,2}=(\frac{3}{2},\pm\frac{\sqrt{3}}{2})$ [and $0=(0,0)$]. We have written $H_{B}$ in a frame rotating with $\omega_{a}$ that we use for the rest of the Hamiltonians of the manuscript. This bath is the bosonic analogue of 
graphene, which can be 
implemented with different systems such 
as cold atoms~\cite{polini13a} or circuit QED \cite{houck12a}. Despite the simplicity of the model, we also expect it to capture the most important features of more complex photonic realizations where Dirac cone dispersions emerge \cite{bravo12a,xie14a}.

\begin{figure}
\centering
\includegraphics[width=0.9\linewidth]{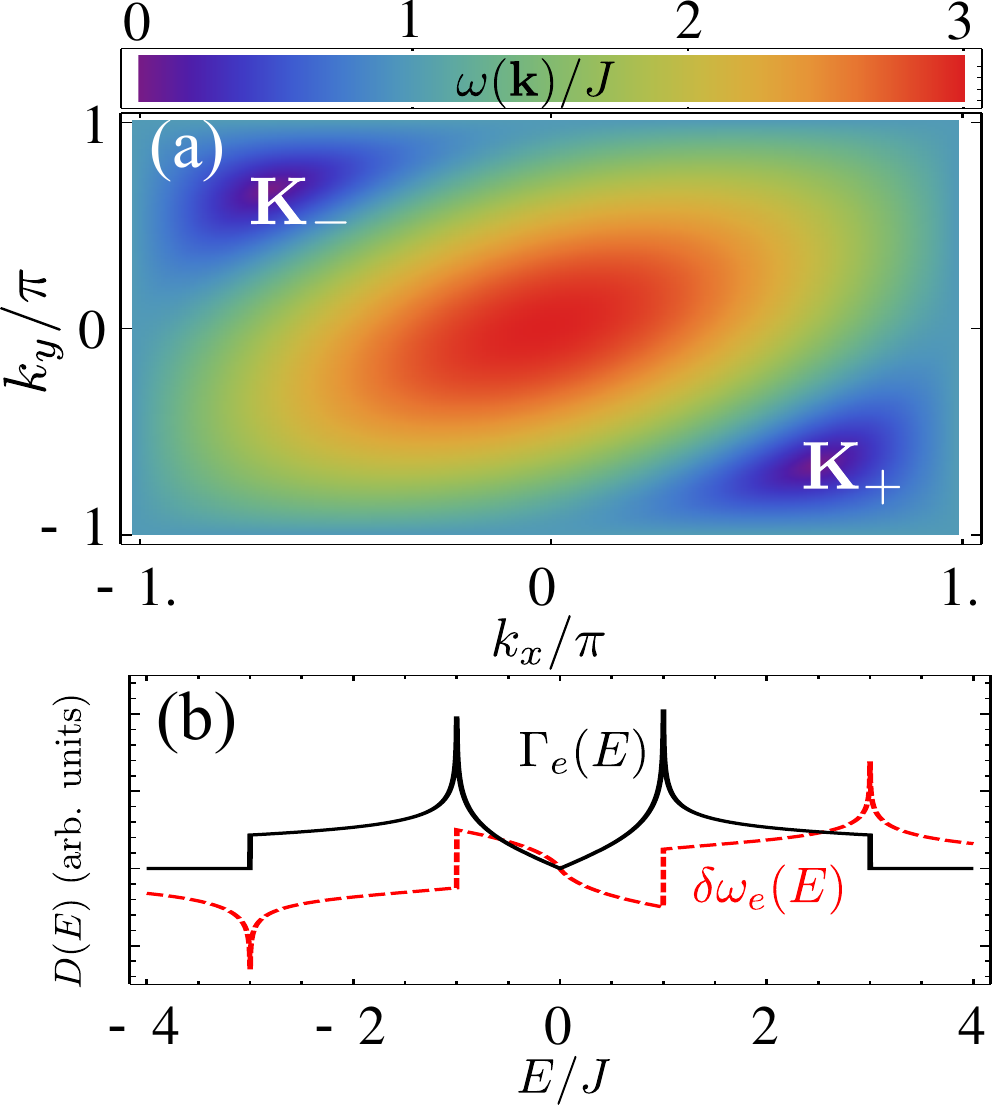}
\caption{(a) Contour plot of $\omega(\kk)$ for the upper band of $H_B$. (b) Density of states $D(E)$ (solid black) for the 2D bath $H_B$, which is proportional to the imaginary part, $\Gamma_e(E)/2$, of the single QE self-energy. For completeness, we plot the real part of the self energy $\delta\omega_e(E)$ (dotted red). }
\label{fig1HC}
\end{figure}

By imposing periodic boundary conditions, we define the operators $\hat c_\kk=\frac{1}{N}\sum_\nn c_\nn e^{-i\kk\cdot\nn}$ for $c=a,b$ bath modes, where $\kk=(k_1,k_2)\equiv k_1 \mathbf{b}_1+k_2\mathbf{b}_2$, with $\mathbf{b}_{1,2}$ satisfying $\aaa_{i}\cdot\bb_{j}=\delta_{ij}$ and $k_{1,2}=\frac{2\pi}{N}(-\frac{N}{2},\dots,\frac{N}{2}-1)$. Using those operators in momentum space, the bath Hamiltonian reads:
\begin{align}
\label{eq:HBBC}
H_B=\sum_\kk \left(f(\kk) \hat{a}^\dagger_\kk \hat{b}_\kk+\mathrm{h.c.}\right)=\sum_\kk \omega(\kk)(\hat{u}^\dagger_\kk \hat{u}_\kk- \hat{l}^\dagger_\kk \hat{l}_\kk)\,,
\end{align}
where the eigenoperators $\hat{u}_\kk(\hat{l}_\kk)=\frac{1}{\sqrt{2}}\left(\hat{a}_\kk+(-)\hat{b}_\kk e^{i\phi(\kk)}\right)$ represent the annihilation operators of upper(lower) band modes, respectively, and $f(\kk)=J(1+e^{i k_1}+e^{i k_2})=\omega(\kk)e^{i \phi(\kk)}$~\footnote{Notice, we use a different convention than the one used in traditional graphene studies~\cite{castroneto09a}.}. 
The dispersion relation extends from $[-3J,3J]$ and has two bands that touch at $\KKK_\pm=\frac{2\pi}{3}(\pm 1,\mp 1)$, the so-called Dirac points corresponding to $\omega(\KKK_\pm)\equiv 0$ [see Fig~\ref{fig1HC}(a)].  At these points, the energy dispersion can be linearized, i.e.,  $f(\KKK_\pm+\qq)\approx J \mathbf{h}_\pm\cdot\qq$, with $\mathbf{h}_\pm=i\left(e^{\frac{\pm 2\pi i}{3}},e^{\frac{\mp 2\pi i}{3}}\right)$. 

The density of states of this bath in the limit $N\rightarrow \infty$ is plotted in Fig~\ref{fig1HC}(b). Apart from discontinuities and divergences also appearing in other structured baths, it possesses a singular point at the Dirac point, i.e., $E=0$. This is the region that we focus along this manuscript as, up our knowledge, has no analogue in other quantum optical scenarios, and it is the source of many interesting behaviour in other contexts~\cite{pereira06a,castroneto09a}.

We are interested in the dynamics of $N_e$ QEs described by two-level systems, $\{\ket{g}_{j,\alpha},\ket{e}_{j,\alpha}\}$, whose free Hamiltonian reads: $H_S=\Delta\sum_{j,\alpha}\sigma_{ee}^{j,\alpha}$, where $\Delta=\omega_e-\omega_a$ is the detuning between the transition frequency of the QEs, $\omega_e$, and the reference energy of the bath modes, $\omega_a$. We also defined the spin operator of the $(j,\alpha)$-th QE as $\sigma_{\alpha\beta}^{j,\alpha}=\ket{\alpha}_{j,\alpha}\bra{\beta}$. The indices $(j,\alpha)$ denote both the position, $\nn_j$, and the sublattice $\alpha=A,B$, of the bath mode that the QE is coupled to. Thus, the interaction Hamiltonian generally reads: 
\begin{equation}
 \label{eq:Hint}
 H_\intt=g\sum_{j,\alpha} \left(d_{\nn_j,\alpha} \sigma_{eg}^{j_,\alpha}+\mathrm{h.c.}\right)\,
\end{equation}
with $d_{j,A/B}=a_\jj/b_\jj$ and $g$ is the coupling strength of the interaction. We study the dynamics of the QEs when they are initialized on a given state $\ket{\Phi_0}_S$ containing a single excitation and the bath is in the vacuum, i.e., $\ket{\mathrm{vac}}_B=\ket{0}_A^{\otimes N^2}\otimes \ket{0}_B^{\otimes N^2}$. Then, we study the evolution of the combined system under the total Hamiltonian $H=H_S+H_B+H_\intt$.
\begin{figure}
\centering
\includegraphics[width=0.8\linewidth]{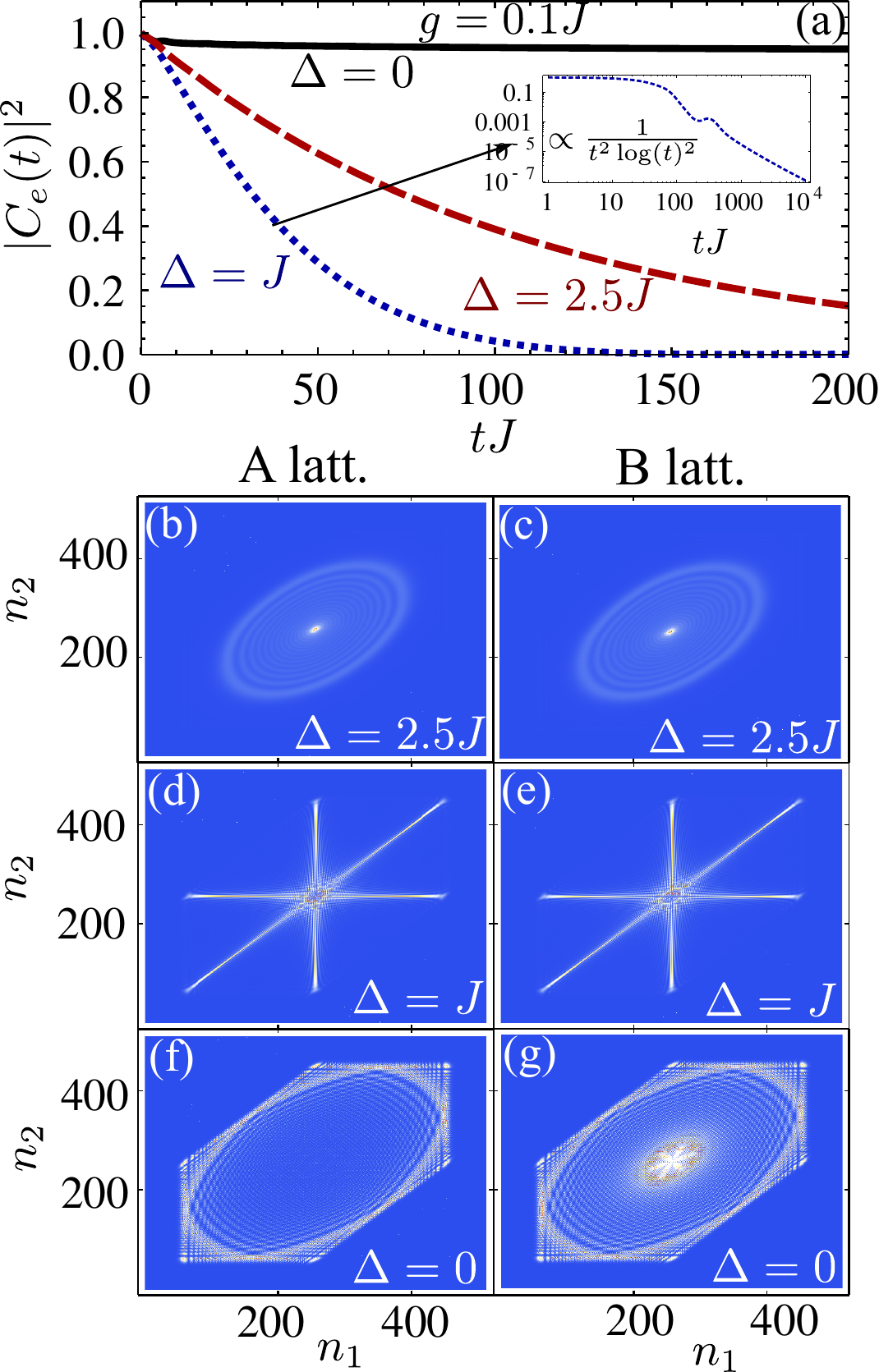}
\caption{(a) Excited state population $|C_e(t)|^2$ of a single QE coupled to a bath with $N=512$ sites in each sublattice with $g=0.1J$ for $\Delta/J=0$ (solid black), $1$ (dotted blue) (also in logarithmic scale at the inset) and $2.5$ (dashed red). (b-e) Corresponding bath probability amplitude in the A/B lattices at $tJ=200$ for the different situations of panel (a) as shown in the legend. Notice that lattice distortion arises from the use of the indices $n_{1,2}$ instead of the real positions $\nn=n_1 \aaa_1+n_2\aaa_2$.}
\label{fig2HC}
\end{figure}

\section{Single quantum emitter dynamics\label{sec:single}}

It is instructive to first study the single excited QE, i.e.,  $\ket{\Phi_0}_S=\ket{e}$, as the structure of the bath has already remarkable consequences on its dynamics. We assume the QE to be coupled to the mode $a_{\nn_0}$ (the coupling to the B lattice leads to similar behaviour). As the total Hamiltonian $H$ conserves the number of excitations, the wavefunction at any time $t>0$ can be written:
\begin{equation}
 \label{eq:wavesingl}
 \ket{\Phi(t)}=\left[C_e(t)\sigma_{eg}+\sum_{\nn_,\alpha=a,b}C_{\nn_,\alpha}(t) d_{\nn_,\alpha}^\dagger  \right]\ket{g}\otimes\ket{\mathrm{vac}}_B\,
\end{equation}
with initial condition $C_e(0)=1$ and $C_{\nn,\alpha}(0)=0$ . In Fig.~\ref{fig2HC} we show the result of a numerical simulation of the dynamics with $g=0.1J$ and for several illustrative $\Delta$'s, together with the bath population in the A/B lattices at $tJ=200$. For $\Delta$'s close to the band edges ($\Delta=2.5J$) the emission is mostly isotropic and equally distributed among the A and B sublattices, as shown in Figs.~\ref{fig2HC}(b-c). The isotropic emission can be traced back to the isotropic character of $\omega(\kk)$ around the band edges. For $\Delta=J$ the emission is highly anisotropic~\cite{mekis99a,langley96a,galve17a,gonzaleztudela17a,gonzaleztudela17b}, predominantly in three directions in both the A/B lattices, as shown in Fig.~\ref{fig2HC}(d-e). This is also accompanied by overdamped oscillations and power law decay dynamics [see inset of Fig.~\ref{fig2HC}(a)]. These non-perturbative dynamics associated to divergences of $D(E)$ were explored in 
Ref.~\cite{gonzaleztudela17a,gonzaleztudela17b}. Finally, for regions close to the Dirac point, the dynamics get slower because of the reduction of the density of modes until it seems to get quenched at $\Delta=0$. This behaviour appears together with an asymmetry in the population of the A/B lattices, as shown in Fig.~\ref{fig2HC}(f-g), where the B lattice shows a larger bath population around the impurity than the A sublattice that the QE is coupled to.  This will have implications when more than one QE are coupled to the lattice, as we show below.

\begin{figure}
\centering
\includegraphics[width=0.8\linewidth]{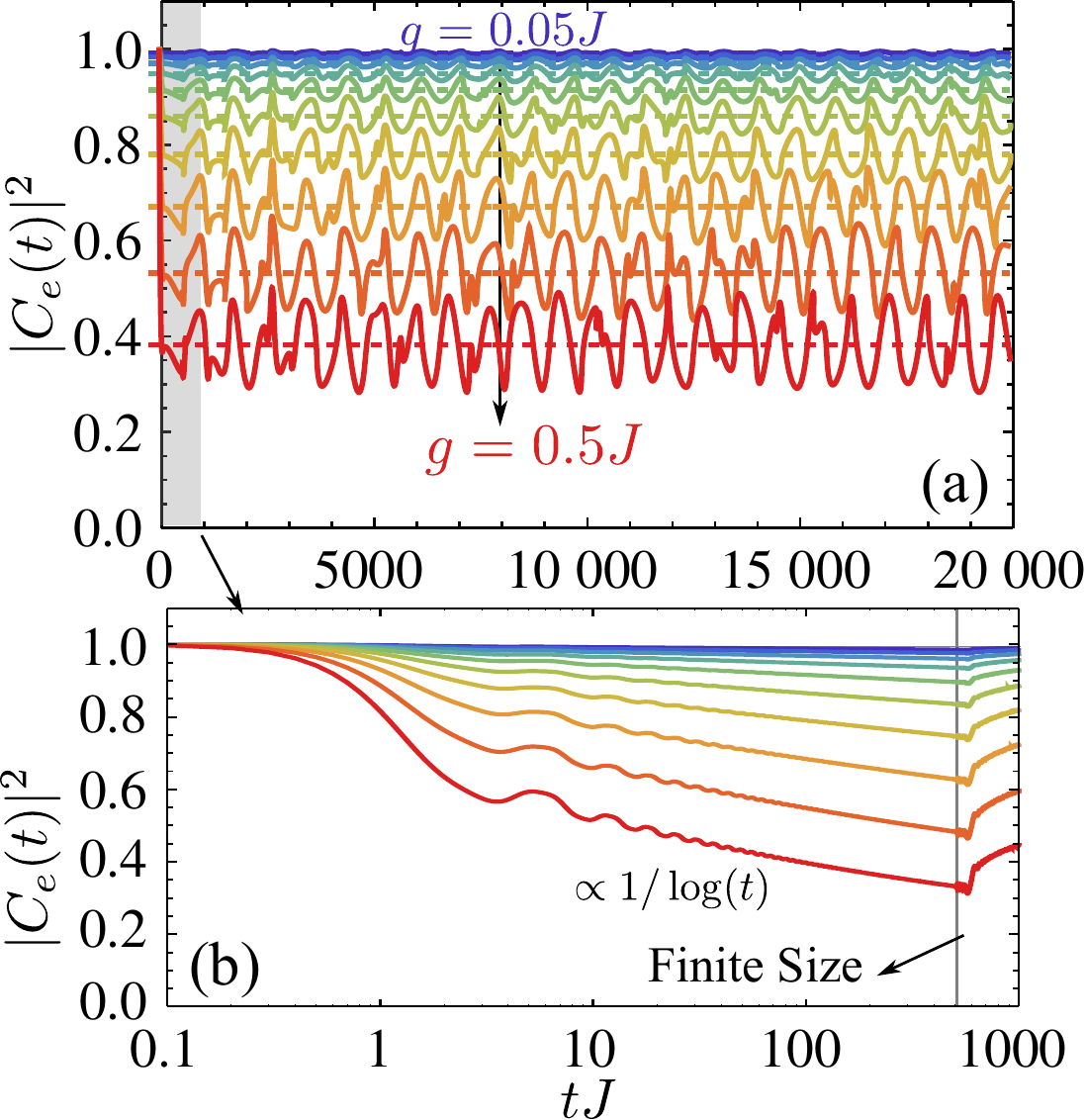}
\caption{a) Excited state population $|C_e(t)|^2$ of a single QE coupled to a bath with $N=512$ with $\Delta=0$ for different logarithmically spaced $g/J$ ranging from $g/J=0.05$ to $0.5$. The dotted lines correspond to the residue of the \emph{quasi} bound state, $|R_0|^2$, as defined in the Eq.~\ref{eqHC:res0}. (b) Same, but zooming in for short times to see the logarithmic decay of excitations.}
\label{fig3HC}
\end{figure}

To gain more intuition about the dynamics at $\Delta=0$, we plot in Fig.~\ref{fig3HC} the excited state population, $|C_e(t)|^2$, for a larger timescale and different ratios $g/J$. We observe that the dynamics at this point have two unexpected behaviours: i) for long times the emission appears to be oscillating around a constant value which is smaller the larger $g/J$; ii) For short times [Fig.~\ref{fig3HC}(b)], $|C_e(t)|^2$ shows a logarithmic relaxation until a time $t_0$ related to the finite size, $N$, of the numerical simulation.

To explain this unconventional behaviour, we use the resolvent operator technique \cite{cohenbook92a}, in which the probability amplitude $C_e(t)$ is calculated as the Fourier transform:
\begin{equation}
\label{eq:cet}
C_e(t)=-\frac{1}{2\pi i}\int_{-\infty}^\infty dE G_e(E+i0^+)e^{-i E t}\,,
\end{equation}
of the single QE Green function, $G_e(z)$. The latter is given by:
\begin{equation}
 \label{eq:probamp}
 G_e(z)=\frac{1}{z-\Delta-\Sigma_e(z)}\,,
\end{equation}
where $\Sigma_e(E)$ is the so-called self-energy which contains the effect that the bath produces in the QE. In contrast to standard single band models, the self-energy contains the contribution of the two baths (upper/lower band), leading to:
\begin{equation}
 \Sigma_e(z)=\frac{g^2}{2N^2}\sum_\kk \left(\frac{1}{z-|f(\kk)|}+\frac{1}{z+|f(\kk)|}\right)=\frac{g^2}{N^2}\sum_\kk \frac{z}{z^2-|f(\kk)|^2}\,.
\end{equation}

This function can be calculated analytically in the limit $N\rightarrow\infty$, in terms of elliptic integrals~\cite{horiguchi72a}, obtaining
\begin{align}
\label{eqHC:self}
 \Sigma_e(z)&=\frac{g^2 z }{4\pi} C(z) \KK^{\mathrm{I}}(k(z)^2)\,, \\
 C(z)&=\frac{8}{(\sqrt{z^2}-J)^{3/2}(\sqrt{z^2}+3J)^{1/2}}\,,\\
 k(z)&=\frac{C(z) \sqrt[4]{z^2}}{2}
\end{align}
for $z\notin[-3,3]J$, and where $\KK^{\mathrm{I}}(m)$ has to be defined in a piecewise manner to guarantee analyticity in its definition domain:
\begin{align}
\KK^{\mathrm{I}}(m)&=\KK(m)\,,\,\mathrm{if}\,\,\mathrm{Im}[k(z)]\mathrm{Im}[z^2]<0\,,\nonumber\\
\KK^{\mathrm{I}}(m)&=\KK(m)+2i K(1-m)\,,\,\mathrm{if}\,\,\mathrm{Im}[k(z)],\mathrm{Im}[z^2]>0\,,\nonumber\\
  \KK^{\mathrm{I}}(m)&=\KK(m)-2i K(1-m)\,,\,\mathrm{if}\,\,\mathrm{Im}[k(z)],\mathrm{Im}[z^2]<0\,,
\end{align}
being $\KK(m)$ the complete elliptic integral of the first kind defined by:
\begin{equation}
\KK(m)=\int_0^{\pi/2}d\theta\frac{1}{\sqrt{1-m\sin^2(\theta)}}\,.
\end{equation}

Evaluating $\Sigma_e(z)$ slightly above the real axis,  $\Sigma_e(E+i0^+)=\delta\omega_e(E)-i\frac{\Gamma_e(E)}{2}$, we obtain the functions $\delta\omega_e(E)$ and $\Gamma_e(E)$, that we plot in Fig.~\ref{fig1HC}(b). There, we observe that the function $\Gamma_e(E)$ [$\delta\omega_e(E)$] has a discontinuity [divergence] at $E=\pm 3J$ appearing at the border of the band edges of the upper/lower band, $\pm \omega(\kk)$. Moreover, at $E=\pm J$ the function also displays a divergence [discontinuity] at $E=\pm J$. Close to the Dirac point, for $|E|\ll J$, the self-energy can be expanded as:
  \begin{align}
  \label{eq:ReeHCexp}
  \Sigma(E+i0^+)\approx \frac{g^2}{\pi\sqrt{3}J^2}\left[E \log\left(\frac{E^2}{9J^2}\right)-i\pi |E|\right]\,,
  \end{align}
such that both $\Gamma_e(E)$ and $\delta\omega_e(E)$ are continuous functions at $E=0$, but have discontinuous derivatives. 

Perturbative approaches simply replace $\Sigma_e(E+i 0^+)\approx \Sigma_e(\Delta+i 0^+)$ in Eq.~\ref{eq:cet}, and assumes that the dynamics of $C_e(t)$ are just given by the contribution of the pole at $z_M=\Delta+\Sigma_e(\Delta+i0^+)=\Delta+\delta\omega_M-i\frac{\Gamma_M}{2}$. Thus, the population dynamics within that approximation are given by $|C_e(t)|^2\approx e^{-\Gamma_M t}$, with $\Gamma_M$ being the one predicted by Fermi's Golden Rule. As $z_M=0$ at $\Delta=0$, this perturbative approach predicts no decay, i.e., $C_e(t)=1$, and fails therefore to describe both the logarithmic relaxation and fractional decay that we observe for finite systems in Fig.~\ref{fig3HC}. 

\begin{figure}
\centering
\includegraphics[width=0.8\linewidth]{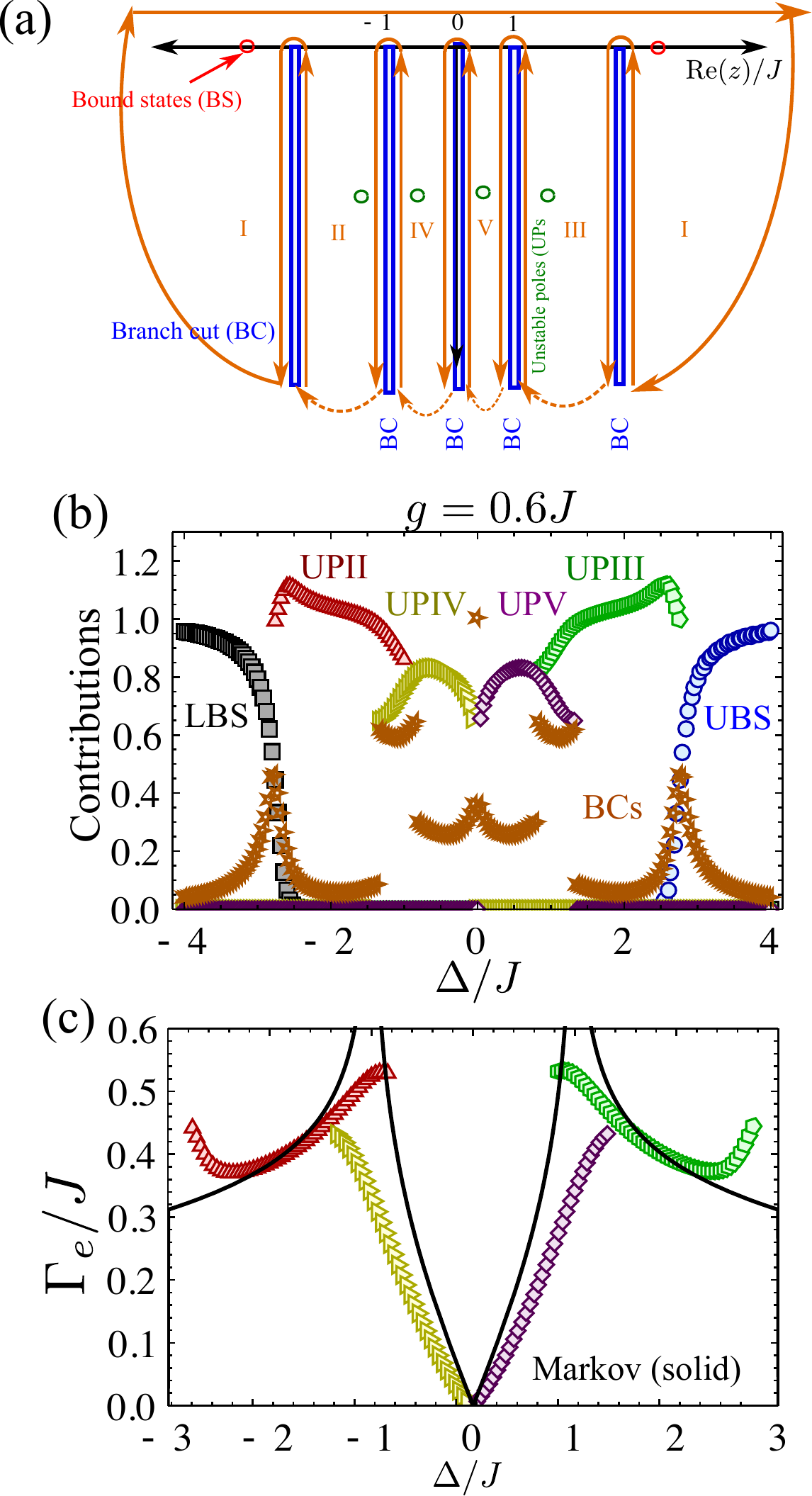}
\caption{(a) Contour of integration to calculate the probability amplitude $C_e(t)$ for the honeycomb bath. The self-energy is discontinuous at $\pm 3J,\pm J,0$ which force to take a detour and that give rise to branch cuts (BCs) contribution to the dynamics. We also denote in red/green the possible real/imaginary poles (BSs/UPs) appearing in the different regions. (b) Absolute value of the different contributions (see legend) of $C_e(t)$ at $t=0$ and $g=0.6J$ as a function of $\Delta$. (c) Comparison between the Markov prediction to the decay rate (solid) and the imaginary part of the UPs numerically obtained by solving the pole equation.}
\label{fig2HCL}
\end{figure}

To obtain the exact dynamics of the probability amplitude $C_e(t)$ one can calculate the Fourier Transform of Eq.~\ref{eq:cet} by closing the contour of integration in the lower plane of the complex plane ($\mathrm{Im}[z]<0$). However, when the $\Sigma_e(z)$ have regions where it is non-analytic, as the one of Eq.~\ref{eqHC:self}, one must be careful when closing the contour to avoid them. One possible choice for the contour of integration consists of taking 5 detours of the integration contour at $E=\pm 3J, \pm J$ and $0$ as schematically depicted in Fig.~\ref{fig2HCL}(a). In the regions II-V, we must analytically continue the function $\Sigma_e(z)$ into other Riemann sheets (see Appendix for details) to ensure the continuity of $\Sigma_e(z)$ along the integration contour. Thus, the total dynamics can be shown to be given by a sum of different contributions:
\begin{equation}
 C_e(t)=R_0+\sum_{jBS} R_{jBS} e^{-i z_{jBS}}+\sum_{jUP} R_{jUP} e^{-i z_{jUP}}+\sum_{\alpha}C_{\alpha BC}(t)\,,
\end{equation}
where we have
\begin{itemize}
\item The contribution from real bound states (BS) of the single QE Green function $G_e(E)$ that appear outside the continuum. Their energies are given by the solution $z_{\BS}-\Delta-\Sigma_e(z_{\BS})=0$, and which corresponding Residue is obtained by using Residue Theorem yielding:
 \begin{equation}
  R_{\BS}=\frac{1}{1-\partial_z\Sigma_e(z)}\Big|_{z=z_\BS}\,.
 \end{equation}
 
\item Apart from the real BS, we can also find complex or unstable poles (UPs) when $\mathrm{Re}[z]\in [-3J,3J]$ appearing in the analytic continuation of the Green Function to other Riemann sheets. Since our Green function is extended into different sheets it is possible to find more than one UP for the same value of $\Delta$. These UPs are obtained from the same pole equation than the real ones, but replacing $\Sigma_e$ by the corresponding analytic continuation of the region. The corresponding residue is also calculated in an analogue way to the BS case.
 
 \item We must also take into account the contributions from the detour because of the BCs. Notice, that in each side of the detour one must use the corresponding $\Sigma_e$ depending on the region where the integral appears. 
 
 \item Finally, we have separated the contribution corresponding to the \emph{quasi-bound state} (qBS) at $z=0$, with residue $R_0$, as it will play an important role in the discussion of finite systems. The residue $R_0$ can be calculated as follows:
\begin{align}
\label{eqHC:res0}
 R_{0}=\left[\frac{1}{1-\partial_z\Sigma_e(z)}\right]_{z=i0^+}=\frac{1}{1+\frac{g^2}{J^2}g(N)}\,,
\end{align}
where $g(N)=\frac{J^2}{N^2}\sum_{\kk}\frac{1}{|f(\kk)|^2}\sim \frac{2}{\pi\sqrt{3}} \log(N)$ for $N\gg 1$ as we show in the Appendix.  This residue can be understood as the overlap of the initial wavefunction with the qBS emerging at $\Delta=0$~\cite{pereira06a}, $\ket{\Phi_{\mathrm{qBS}}}$, i.e., $R_0=\braket{\Phi_0}{\Phi_{\mathrm{qBS}}}$ and is ultimately responsible of the excitation remaining in the QE as $C_e(\infty)=R_0$. For a finite system, the qBS is indeed square integrable which is why $R_0$ can be finite. However, the very slow decay ($1/r$) with the distance makes it not square integrable in the thermodynamic limit, which is why $R_0\rightarrow 0$ when $N\rightarrow \infty$.
 
 \item The rest of the contours can be shown to give no contribution.
\end{itemize}

To illustrate the weight in the thermodynamic limit of the different processes to the dynamics depending on $\Delta$, we plot the absolute value of each contribution at $t=0$ for fixed $g=0.6J$ and different $\Delta$'s in Fig.~\ref{fig2HCL}(b). For the sake of simplicity, we plot the sum of all the branch cut contributions using the same color, although we calculate them independently. We observe several interesting regimes:
\begin{itemize}
 \item When $\Delta$ is tuned very far from the band, the main contributions is either from the LBS or UBS. This predicts that the QE will not decay as there are no modes energetically resonant with the QE. As expected from the divergence of $\delta\omega_e(E)$ at $E=\pm 3J$~\cite{shi16a}, both the UBS/LBS survive for all $\Delta$, including those which lie within the band.
 \item As the QE transition frequency gets closer to the band edges, $|\Delta\pm 3J|\ll g,J$, the contribution of the BS decreases and the BC contribution becomes more important, until the weight moves to the UPs.
 \item As we already showed for a square lattice~\cite{gonzaleztudela17a,gonzaleztudela17b}, the divergence of the imaginary part of $\Sigma_e(E)$, in this case at $E=\pm J$, leads to the coexistence of two UPs for a range of $\Delta$'s given by the jump of $\delta\omega_e(E)$ around that value. Moreover, the imaginary part of the pole saturates to a finite value, as shown in Fig.~\ref{fig2HCL}(c), compared to the infinite value predicted by Fermi's Golden Rule. As we already explored extensively this phenomenon in Refs.~\cite{gonzaleztudela17a,gonzaleztudela17b}, we will not investigate it further in this manuscript.
 \item Finally, around the Dirac point we observe that as $\Delta$ gets closer to $0$ the BC contribution increases, making a discontinuous jump at $\Delta=0$, where the contribution of the left/right UPs vanishes. The middle BC contribution can be calculated by using the analytical expansions of the elliptic functions around $E\approx 0$ which yield:
\begin{equation}
 \lim_{t\rightarrow\infty} C_{\mathrm{MBC}}(t)\approx  -\frac{\pi \sqrt{3}J^2}{2 g^2} \int_0^\infty dy e^{-y t} \frac{1}{y\log\left(\frac{y}{3J}\right)^2}\,.
\end{equation}

The asymptotic expansion of Laplace transform of functions with logarithmic singularities was studied in Ref.~\cite{wong78a}, where it was shown that:
\begin{align}
\label{eqHCL:sing}
& \lim_{t\rightarrow\infty}\int_0^\infty dy e^{-yt} y^{\alpha-1} (-\log(y))^\beta =\nonumber \\&=\frac{1}{t^\alpha}\sum_{k=0}^\infty (-1)^k\binom{\beta}{k}\Gamma^{(k)}(\alpha)\big(\log(t))^{\beta-k}\,.
\end{align}
where $\mathrm{Re}\alpha$ must be $>0$. Unfortunately, our $C_{\mathrm{MBC}}(t)$ lies at the border of validity of this formula as $\alpha=0$. However, one can realize that by taking the first derivative of the integral once that we obtain a formula that lies within the region of validity:
\begin{equation}
\frac{d}{dt}\left[\int_0^\infty dy e^{-yt} y^{-1} (\log(y))^{-2}\right]=-\int_0^\infty dy e^{-yt}  (\log(y))^{-2}\,,
\end{equation}
which has the shape of Eq.~\ref{eqHCL:sing}, but with $\alpha=1$. Applying now the formula, the leading term of the derivative at long times is $\sim 1/(t\log(t)^2)$. Now, integrating back this leading term one can find:
\begin{equation}
\label{eq:CMBClead}
\lim_{t->\infty} C_{\mathrm{MBC}}(t)\approx  -\frac{\pi \sqrt{3}J^2}{2 g^2\log(3 J t)}\,.
\end{equation}
such that $|C_{\mathrm{MBC}}|^2\propto 1/(\log(t))^2$. We also checked numerically that the convergence to the leading term is slow such that intermediate times one may find other scalings for the relaxation.
\end{itemize}

Summarizing, for a single QE we have shown that in the vicinity of the Dirac point, perturbative methods fail to capture the exact behaviour of both finite and infinite systems. In finite systems, it does not capture the existence of fractional decay of excitations ($R_0<1$), whereas in infinite ones it completely neglects the logarithmic relaxation which dominates the dynamics.

\section{Two quantum emitters dynamics\label{sec:two}}

Now we will study the consequences of such exotic behaviour when two QEs are coupled to the bath. In particular, we aim at discerning whether it is possible to obtain long-range  dipole-dipole interactions by exploiting the existence of the qBS. To explore that situation, we assume that two QEs initially start in $\ket{\Psi_0}_S=\ket{e}_1\otimes\ket{g}_2$ and study the populations of the QEs, $|C_{1,2}(t)|^2$, to see whether there are coherent oscillations or not. 

\subsection{Rewriting interaction Hamiltonian for two QEs}

Before moving on to the solution of the dynamics $|C_{1,2}(t)|^2$ it is instructive for this Section, to rewrite $H_\intt$ in terms of the eigenmodes of the $H_B$ operator, which reads:
\begin{align}
 \label{eqHCL:Ham}
  H_\intt&=\frac{g}{N\sqrt{2}}\sum_{j}\sum_\kk\left( \hat u_{\kk} \left( e^{-i\kk\cdot\nn_j} \sigma_{eg}^{j,A}+e^{-i(\kk\cdot \nn_j+\phi(\kk))}\sigma_{eg}^{j,B}\right)+\mathrm{h.c.}\right)+\nonumber\\
  &+\frac{g}{N\sqrt{2}}\sum_{j}\sum_\kk\left(\hat l_{\kk} \left( e^{-i\kk\cdot\nn_j} \sigma_{eg}^{j,A}-e^{-i(\kk\cdot\nn_j+\phi(\kk))}\sigma_{eg}^{
 j,B}\right)+\mathrm{h.c.}\right)\,.
\end{align}

Let us particularize for the two situations of interest of this Section, namely,
\begin{itemize}
 
 \item  For two QEs at positions $\nn_{1,2}$ coupled within the same sublattice, e.g., A, the interaction Hamiltonian reads:
 \begin{align}
  H_\intt=\frac{g}{\sqrt{2} N} \sum_{j} \sum_\kk \left[ e^{-i\kk\cdot \nn_j}\sigma_{eg}^j \left(\hat u_\kk+\hat l_\kk\right)+\mathrm{h.c.}\right]\,,
 \end{align}
 where we dropped the index $\alpha$ denoting the sublattice the QE is coupling to.  Apart from the coupling of the QEs to two independent baths (upper/lower modes), the only difference between the coupling of the two QEs to each bath is the phase difference introduced by $e^{i\kk \cdot \nn_{12}}$, where we use the notation $\nn_{12}=\nn_2-\nn_1$ for the vector connecting the two impurities. This is analogue to what happens with standard open quantum systems.
  
 \item Different from the previous situation, when two QEs at positions $\nn_{1,2}$ are coupled to \emph{different} sublattices, e.g., the $\nn_{1/2}$ one to be coupled the A/B lattice respectively. In that case, the interaction Hamiltonian is given:
 \begin{align}
 \label{eqHCL:Ham2}
  H_\intt&=\frac{g}{N\sqrt{2}}\sum_\kk\Big[\left(\hat u_{\kk} +\hat l_{\kk} \right) e^{-i\kk\cdot\nn_1} \sigma_{eg}^{1}+\nonumber \\
  &\left( \hat u_{\kk} -\hat l_{\kk} \right)e^{-i(\kk\cdot\nn_2+\phi(\kk))}\sigma_{eg}^{2}+\mathrm{h.c.}\Big]\,.
\end{align}
 which apart from the propagation phase ($e^{i\kk\cdot\nn_{12}}$) contains an extra $\kk$-dependent phase, $e^{i\phi(\kk)}$, that has consequences on the dynamics.
 \end{itemize}

 It is also interesting to consider the bath modes that are coupled to the QE symmetric/antisymmetric operators $\sigma^\dagger_\pm=\frac{1}{\sqrt{2}}\left(\sigma_{eg}^{1}\pm \sigma_{eg}^{2}\right)$ and $\sigma_{\pm}=\left(\sigma_{\pm}^\dagger\right)^\dagger$. With that definition, it is possible to rewrite $H_{\mathrm{int}}$ in such a way that $\sigma^\dagger_\pm$ couple to two orthogonal bath modes for the upper/lower band operators. This allows us to solve the symmetric/antisymmetric dynamics independently.  In particular, for the case of two QEs coupled to the AA lattice, $H_\mathrm{int}$ is rewritten as:
\begin{align}
H^{\mathrm{AA}}_\intt &=\frac{g}{N}\sum_{\kk>0,\pm}\left[\sqrt{1\pm\cos(\kk\cdot\nn_{12})}\left(\hat u_{\kk,\pm}+\hat l_{\kk,\pm}\right)\sigma_{\pm}^\dagger+\mathrm{h.c.}\right]\,,
\end{align}
where $u_{\kk,\pm},l_{\kk,\pm}$ are two orthogonal bath modes defined as:
\begin{align}
\hat u_{\kk,\pm}&=\frac{1}{2\sqrt{1\pm\cos(\kk\cdot\nn_{12})}}\Big[\left(e^{-i\kk \cdot \nn_1}\pm e^{-i\kk \cdot \nn_2}\right)\hat u_{\kk}+\nonumber\\
&\left(e^{i\kk \cdot \nn_1}\pm e^{i\kk \cdot \nn_2}\right)\hat u_{-\kk}\Big]\,,\\
\hat l_{\kk,\pm}&=\frac{1}{2\sqrt{1\pm\cos(\kk\cdot\nn_{12})}}\Big[\left(e^{-i\kk \cdot \nn_1}\pm e^{-i\kk \cdot \nn_2}\right)\hat l_{\kk}+\nonumber\\ &\left(e^{i\kk \cdot \nn_1}\pm e^{i\kk \cdot \nn_2}\right)\hat l_{-\kk}\Big]\,.
\end{align}

For the case of two QEs, $\nn_1$ and $\nn_2$, coupled to the A/B lattice respectively, the interaction Hamiltonian, $H_\intt$, changes to:
\begin{align}
H^{\mathrm{AB}}_\intt&=\frac{g}{N}\sum_{\kk>0,\pm}\Big[\sqrt{1\pm\cos(\kk\cdot\nn_{12}+\phi(\kk))}\hat u_{\kk,\pm}\sigma_{\pm}^\dagger+\nonumber\\ 
&\sqrt{1\mp \cos(\kk\cdot\nn_{12}+\phi(\kk))}\hat l_{\kk,\pm}\sigma_{\pm}^\dagger+\mathrm{h.c.}\Big]\,,
\end{align}
where $\hat u_{\kk,\pm},\hat l_{\kk,\pm}$ are two orthogonal bath modes defined as:
\begin{align}
\hat u_{\kk,\pm}=\frac{1}{2\sqrt{1\pm\cos(\kk\cdot\nn_{12}+\phi(\kk))}}&\Big[\left(e^{-i\kk \cdot \nn_1}\pm e^{-i(\kk \cdot \nn_2+\phi(\kk))}\right)\hat u_{\kk}+\nonumber\\
&\left(e^{i\kk \cdot \nn_1}\pm e^{i(\kk \cdot \nn_2+\phi(\kk))}\right)\hat u_{-\kk}\Big]\,,\\
\hat l_{\kk,\pm}=\frac{1}{2\sqrt{1\mp \cos(\kk\cdot\nn_{12}+\phi(\kk))}}&\Big[\left(e^{-i\kk \cdot \nn_1}\mp e^{-i(\kk \cdot \nn_2+\phi(\kk))}\right)\hat l_{\kk}+\nonumber\\ &\left(e^{i\kk \cdot \nn_1}\mp e^{i(\kk \cdot \nn_2+\phi(\kk))}\right)\hat l_{-\kk}\Big]\,.
\end{align}

The important consequence of this separation is that for the case of two QEs we will be able to work separately with the symmetric \& antisymmetric subspaces. This means that the Green functions $G_{1,2}(z)$ associated to the probability amplitudes $C_{1,2}(t)$ can be obtained from Green Functions $G_{\pm}(z)$ associated to the symmetric/antisymmetric combination of excitations as follows: $G_{1,2}(z)=\left[G_{+}(z)\pm G_{-}(z)\right]/2$. Remarkably, $G_{\pm}(z)$ are given by the same expression than the single QE, $G_e(z)$, just replacing $\Sigma_e(z)\rightarrow \Sigma^\beta_{\pm}(z;\nn_{12})=\Sigma_e(z)\pm \Sigma^\beta_{12}(z;\nn_{12})$. The collective self-energy, $\Sigma^\beta_{12}(z;\nn_{12})$, reads:
\begin{equation}
\label{eqHC:Self12}
 \Sigma^\beta_{12}(z;\nn_{12})=\frac{g^2}{N^2}\sum_\kk \frac{D_\beta e^{i\kk\cdot \nn_{12}}}{z^2-|f(\kk)|^2}\,,
\end{equation}
where $\beta=AA,AB,BA,BB$ is an index that denotes to which sublattices the QEs are coupled to. Depending on $\beta$, we have that: $D_{AA}=D_{BB}=z$, whereas $D_{AB}=D_{BA}^*=f^*(\kk)$.

Perturbative approaches predict a dynamics in the symmetric and antisymmetric subspace $C_{\pm}(t)\approx e^{-i(J_{M,\pm}-i\frac{\Gamma_{M,\pm}}{2})t}$, where we define $J_{M,\pm}-i\frac{\Gamma_{M,\pm}}{2}=\Sigma_{\pm}(\Delta+i0^+)$. Thus, in the original basis they turn into:
\begin{align}
 \label{eqHC:c12}
 |C_{1,2}(t)|^2&\approx \frac{1}{4}\Big[ \pm 2 e^{-\frac{\Gamma_{M,+}+\Gamma_{M,-}}{2}t}\cos((J_{M,+}-J_{M,-})t)\nonumber \\
 & +e^{-\Gamma_{M,+} t}+e^{-\Gamma_{M,-} t}\Big]\,,
\end{align}

Therefore, purely coherent oscillations of $|C_{1,2}(t)|^2$ are associated to the existence of two real bound states[$\Gamma_{M,\pm}=0$], $\ket{\Psi_{\pm}}$, of the symmetric/antisymmetric component with different real component, $J_{M,+}-J_{M,-}\neq 0$.

\subsection{Two QEs coupled to the AA/BB lattices}

At this point, it is easy to show that $\Sigma_{\pm}^{\mathrm{AA/BB}}(0)=0$, such that $z=0$ satisfies the pole equation for both the symmetric and antisymmetric wavefunction. Within a perturbative treatment, this will imply that these states are subradiant and should not relax to the ground state. However,  as it occurred for a single QE when doing the exact calculation one finds different result. In particular, it can be shown that the residue associated to $z=0$ reads:
\begin{equation}
 R_{\pm}^{\mathrm{AA/BB}}=\frac{1}{1+\frac{g^2}{J^2} g_{\pm}(N)}\,,
\end{equation}
where:
\begin{equation}
\label{eqHCL:sumgpm}
 g_{\pm}(\nn_{12},N)=\frac{J^2}{N^2}\sum_\kk \frac{1\pm e^{i\kk \nn_{12}}}{|f(\kk)|^2}\,.
\end{equation}

Reminding that the divergence with $N$ occurs because $f(\KKK_{\pm})=0$ at the $\KKK_{\pm}$ points, we see that there are two interesting regimes depending on the relative position between the QEs, $\nn_{12}=(n_1,n_2)$.
\begin{itemize}
 \item When $\nn_{12}$ is such that $e^{i\KKK_{\pm}\cdot \nn_{12}}\neq 1$, then the numerator is always finite at the $\KKK_\pm$ points, yielding $g_{\pm}(\nn_{12},N)\propto \log(N)$. The dynamics in the symmetric/antisymmetric subspaces is then very similar to the single QE situation, that is, an initial logarithmic decay for short times, until it quenches for a finite system or completely decays for an infinite one.
 \item On the other hand when $e^{i\KKK_{\pm}\cdot \nn_{12}}= 1$, which occurs when $n_1-n_2=3m$, with $m\in\mathbb{Z}$, then $g_{-}(\nn_{12},N)\rightarrow C(\nn_{12})$ when $N\rightarrow \infty$. This occurs because the divergence of the denominator cancels with the zero of the numerator. This means that for such positions, the antisymmetric component is indeed a real pole of the Green Function in the continuum limit, that is, a perfect subradiant state for all parameter regimes. As it occurs with subradiant states in other reservoirs, the only effect of the distance is to decrease the residue of the pole due to retardation effects. We can estimate these retardation effects, for example, for the situation $n_1=n_2=n$ by assuming that most of the contribution is coming from the $\kk$ modes around $\KKK_{\pm}$, where the divergence of the denominator occurs. Then, we can displace the integral to $\kk=\KKK_{\pm}+\qq$ to obtain:
 \begin{align}
  &\frac{2}{(2\pi)^2}\iint d^2\qq \frac{1-e^{i(q_1+q_2)n}}{q_1^2+q_2^2-q_1 q_2}=\frac{1}{ \pi^2\sqrt{3}}\iint d^2\pp \frac{1-e^{i 3 p_1 n}}{|\pp|^2}=\nonumber\\
  &=\frac{2}{ \pi\sqrt{3}}\int_0^{q_c} dp\frac{1-J_0(3 |\pp| n)}{|\pp|}\approx D+\frac{2}{ \pi\sqrt{3}}\log(n)
 \end{align}
 where the $\pp$-coordinates are defined by $q_{1,2}=\frac{3}{2}\left(p_1\pm \frac{1}{\sqrt{3}} p_2\right)$ and they lead to an isotropic $\omega(\pp)$. In the last approximation of the integral we expanded for $n\gg 1$. The constant $D$ depends on the momentum cut-off $q_c$ and we numerically calculate it using the exact sum of Eq.~\ref{eqHCL:sumgpm} for $\nn_{12}=(1,1)$ obtaining $D\approx 0.6$. Thus, the final residue of these subradiant states is given by:
 \begin{equation}
  R_{\mathrm{sb}}^{AA,BB}\approx \frac{1}{1+\frac{g^2}{J^2}\left(0.6+\frac{2}{\sqrt{3}\pi}\log(n)\right)}\,.
 \end{equation}

 Notice, the very slow decay of the residue of such subradiant states with $n$, i.e., $R_\mathrm{sb}\propto 1/\log(n)$, which is considerably slower than the one of the subradiant states for the square lattice~\cite{gonzaleztudela17a,gonzaleztudela17b} which decay with $1/n^2$ or even slower than the 1D situation $\propto 1/n$.
\end{itemize}

Even though these subradiant states are interesting on its own, their associated real part, which is the one responsible for coherent interactions, is also zero. As in this manuscript we are interested in obtaining coherent interactions, we will not discuss them further, and leave for future work a more detailed study about them.

\subsection{Two QEs coupled to different sublattices.}

\begin{figure}
\centering
\includegraphics[width=0.8\linewidth]{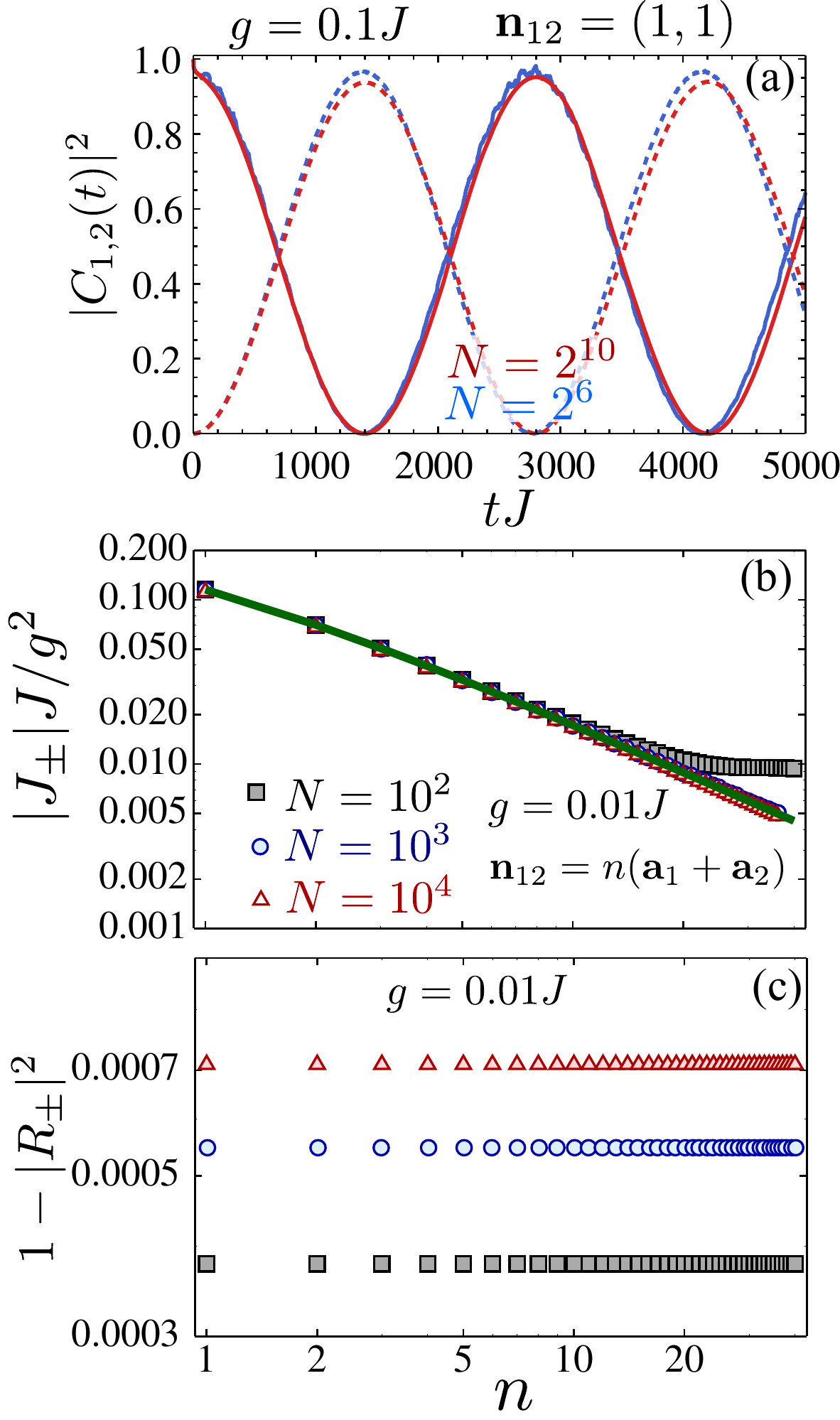}
\caption{a) Excited state population of two QEs separated by $\nn_{12}=(1,1)$ with $\Delta=0$ and coupled to a bath of size $N=2^6$ (blue) and $N=2^{10}$ (red) with $g=0.1J$. (b-c) Scaling of dipole-dipole coupling obtained from the exact pole equation (and the corresponding residue) for two QEs with relative position $\nn_{12}=n(\aaa_1+\aaa_2)$  and coupling $g/J=0.01$ as a function of $n$ and for different $N=10^2,10^3,10^4$. In solid green we plot the corresponding Markov prediction from Eq.~\ref{eqHC:Self12} with $z=0$.}
\label{fig4HC}
\end{figure}

The situation with two QEs coupled to different sublattices is radically different from the single QE or when the two QEs are coupled to the same sublattice. To illustrate it, we plot in Fig.~\ref{fig4HC}(a), the numerical evolution of $C_{1,2}(t)$ for two QEs coupled to the A/B lattice respectively,  with $g=0.1J$, for bath sizes of $N=2^6$ (blue) and $2^{10}$ (red), and relative position $\nn_{12}=\nn^A_{1}-\nn^B_{2}=(1,1)$. We observe (almost) complete oscillations with a very slight dependence of both the frequency and amplitude on the system size, $N$. 

Let us first try to explain these numerical results through perturbative approaches. In this case, the collective self-energy within the Born-Markov approximation is finite and real in the limit $N\rightarrow \infty$, i.e.,  $\Sigma^{AB}_{\pm}(i0^+;\nn_{12})=J_{M,\pm}(\nn_{12})=\pm J_{AB,M}(\nn_{12})\in\mathbb{R}$, which predicts therefore coherent oscillations. We use Eq.~\ref{eqHC:Self12} with $z=0$ to calculate numerically $J_{AB,M}(\nn_{12})$ that we plot in solid green in Fig.~\ref{fig4HC}(b). An asymptotic expression of $J_{AB,M}(\nn_{12})$ for large distances $|\nn_{12}|\gg 1$ reads~\cite{basko08a,bena09a} (see Appendix):
 \begin{align}
\label{eq:selfAABM}
&J_{AB,M}(\nn_{12})=\frac{g^2}{J}\frac{\sqrt{3}}{\pi |\mm_{12}|}\times\nonumber\\ &\times\frac{m_1\cos\left(\frac{2\pi}{3}\left(n_1-n_2\right)\right))-m_2\sin\left(\frac{2\pi}{3}\left(n_1-n_2\right)\right))}{|\mm_{12}|}\,.
\end{align}
where $\mm_{12}=(m_1,m_2)=\left(\frac{3}{2}\left(n_1+n_2\right),\frac{\sqrt{3}}{2}\left(n_1-n_2\right)\right)$ is a rescaled vector in terms of the original coordinates of $\nn_{12}=(n_1,n_2)$. For the situation we are plotting in Fig.~\ref{fig4HC}, i.e., $\nn_{12}=(n,n)$, reads:
\begin{equation}
\label{eqHC:Markov12}
J_{AB,M}(\nn_{12})\approx \frac{g^2}{J}\frac{1}{\pi \sqrt{3} n}\propto \frac{1}{n}\,.
\end{equation}

Summing up, perturbative approaches predict the QEs should experience purely coherent interactions even in the limit $N\rightarrow \infty$. However, as we show in the single QE situation, the perturbative approach may introduce artifacts at points where the self-energy is not analytical. Thus, it is relevant to analyze to what extent the previously mentioned results survive after an exact analysis. 

When we take into account the analytical structure of $\Sigma_{\pm}^\beta$, we find that the dynamics in the symmetric/antisymmetric subspace is given by:
\begin{align}
 C_{\pm}(t)=R_{\pm}e^{-i(J_{\pm}-i\frac{\Gamma_{\pm}}{2})t}+C_{\pm,\mathrm{MBC}}(t)+\mathrm{others}\,,
\end{align}
where $R_{\pm}$ is the residue associated to the pole of $G_{\pm}(z)$ at $z_{\pm}=J_{\pm}-i\frac{\Gamma_{\pm}}{2}$, whereas $C_{\pm,\mathrm{MBC}}(t)$ is the contribution of the detours associated to the branch cut in the middle of the band. The rest of the contributions can be shown to be small for $g/J\ll 1$. In Fig.~\ref{fig4HC}(b), we plot the results obtained by solving the exact pole equation for $N=10^2,10^3$ and $10^4$, showing that indeed the coherent dipole-dipole interactions, $z_{\pm}=\pm J_{AB}\in\mathbb{R}$, survive even for large systems. Interestingly, one can show that the exact dipole-dipole interactions now depend on the system size, $N$, renormalizing to:
\begin{equation}
\label{eqHC:Markov12}
J_{AB}(\nn_{12};N)\approx R_0(N) J_{AB,M}(\nn_{12})\,.
\end{equation}

This expression has a very intuitive meaning: the qBS is able to mediate coherent interactions as long as the overlap with a QE excitation, $R_0(N)$, is finite. In the thermodynamic limit, these coherent interactions vanish as $R_0=0$. However, they decrease so slowly with the system size, $J_{AB}(\nn_{12},N)\propto 1/\log(N)$, that in practical situations they can be harnessed even for very large systems, as shown in Fig.~\ref{fig4HC}(a-b). Another physical picture to understand the decrease of the interactions with the system size is that the mode volume of the qBS, which is the one mediating the interactions, grows with the system size. As it occurs in cavity QED~\cite{ritsch13a}, the strength of the light-mediated interactions is inversely proportional to the mode volume of the photon mediating the interactions. As the mode volume for the qBS is infinite in the thermodynamic limit, the mediated interactions are strictly zero in that limit.

The last remaining thing to prove is that the associated residue for the exact pole, i.e., the overlap with the initial wavefunction, is finite. We calculate it numerically using the expression:
\begin{align}
 \label{eqHC:respm}
 R_\pm=\braket{\Psi_{\pm}}{\Psi_0}_S=\left[\frac{1}{1-\partial_z\Sigma^{AB}_{\pm}(z;\nn_{12})}\right]_{z=J_{M,\pm}}\,.
\end{align}
and show the results in Fig.~\ref{fig4HC}(c). Moreover, we are also able to obtain an asymptotic expansion of $R_\pm$ in the limit $g/J\ll 1$, showing that $R_{\pm}\approx R_0$ (see Appendix), which tell us that the overlap can be very large as long as $\frac{g^2}{J^2}\log(N)\ll 1$. To sum up, the perturbative results for two QEs coupled to the AB lattices get corrected in two ways: i) the dipole-dipole interactions depend on the system size as $J_{AB}(\nn_{12};N)\propto J_{AB,M}(\nn_{12})/\log(N)$, and therefore vanish in the limit $N\rightarrow \infty$; ii) the residues of the associated poles also depend on the system size $R_\pm\propto 1/\log(N)$, also vanishing in the thermodynamic limit.

Finally, we predict that the interaction of QEs of baths with Dirac Cone dispersion relations leads to very exotic models if one considers many QEs: apart from the long-range character of the interactions, they occur only within different sublattices. The predicted many-body spin Hamiltonian can be written as:
\begin{align}
H_{\mathrm{eff}}=\sum_{\nn_A,\mm_B} J_{AB}(\nn_A-\mm_B;N)\sigma^{\nn_A}_{eg}\sigma^{\mm_B}_{ge}
\end{align}
where $\nn_A,\mm_B$ denotes that the coupling is only induced between the QEs coupled to different sublattices and $J_{AB}$ decreases very slowly with the system size, as $1/\log(N)$. This will certainly give rise to very rich many-body behavior without analogue in previously studied models.

\section{Experimental considerations~\label{sec:experiment}}

Along this paper we have discussed a very general model, which can be implemented in very different platforms, ranging from atoms coupled to the guided photons in photonic crystal structures~\cite{bravo12a,xie14a} or even purely atomic implementations using cold atoms trapped in state-dependent optical lattices~\cite{devega08a,navarretebenlloch11a}. In the latter, two atomic states must be trapped in deep/shallow potentials to play the role of the QE/photons respectively. This can be achieved, for example, with Alkaline-Earth atoms in which an optically excited state exists with very long lifetime (of the order of seconds~\cite{schreiber15a,daley08a,snigirev17a}) which allows one to design very different trapping potentials for ground/excited states. This is a very attractive platform because it combines: i) large flexibility in the design of the trapping potentials, and therefore, of the bath geometry; ii) low decoherence rates $\sim$ Hz~\cite{bloch08a,schreiber15a,daley08a,snigirev17a} compared to the 
tunneling  ($J$) and coupling 
($g$) rates which can be of the order of $10$ KHz~\cite{bloch08a}; iii) the possibility of single-site addressing and detection~\cite{sherson10a,weitenberg11a}. 

Irrespective of the realization, the coupling to external reservoirs will induce extra decoherence in the bath/QE modes at rates $\kappa/\Gamma^*$ respectively. To illustrate the effect of the losses on the observed phenomena, we will consider a particular situation where the solution to the dynamics can be obtained analytically. As was shown in Ref.~\cite{gonzaleztudela17b} when $\kappa=\Gamma^*\equiv \Gamma_\mathrm{loss}$, the solution of the density matrix of the combined system of the QE and bath is given by:
\begin{align}
\rho(t)&=e^{-\Gamma_\mathrm{loss} t} \ket{\Psi(t)}\bra{\Psi(t)}+\nonumber \\
&(1-e^{-\Gamma_\mathrm{loss} t})\ket{g\dots g}_S\bra{g\dots g}_S\otimes\ket{\mathrm{vac}}_B\bra{\mathrm{vac}}_B\,.
\end{align}
where $\ket{\Psi(t)}=e^{-i H t}\ket{\Psi(0)}$, i.e., the evolved state in the absence of losses. Thus, the main effect of the losses is to exponentially attenuate with time of the dynamics induced by the system bath coupling in $\ket{\Psi(t)}$. With that solution it is possible to estimate the conditions to observe the emergent phenomena described in this paper.

Observing the complete logarithmic relaxation will be challenging due to its slow decay. However, by going to situations where $g\sim O(J)$, which can be obtained in the cold atoms realization, one can make the dynamics faster. For example, in Fig.~\ref{fig3HC} we see how for $g=0.5J$, the QE has relaxed to approx. 20\% from its initial value at times $tJ\approx 1000$, which will be within the reach of state-of-the-art parameters for the cold atom setup. Regarding the oscillations between two QEs at a given distance, $R$, they will be visible as long as $J_{AB}(\nn_{12})/\Gamma_{\mathrm{loss}}\sim \frac{g^{2}}{J R}\gg 1$. In that case, the finite lifetime $\Gamma_{\mathrm{loss}}^{-1}$ sets a critical distance where these oscillations can still be observed.

\section{Conclusions \label{sec:conclu}}

 To sum up, we have studied the quantum dynamics emerging when 
QEs are interacting though a reservoir with hexagonal lattice symmetry, focusing on the physics emerging because of the Dirac cone dispersion relation of the bath modes. For a single QE, we show that an exact treatment predicts the fractional decay of excitations and a logarithmic relaxation which can not be captured by perturbative treatments. Moreover, when several QEs are coupled to the reservoir purely long range coherent dipole interactions emerge between QEs coupled to different sublattices, which leads to very exotic many-body spin Hamiltonians. Further flexibility to engineer the QEs coupling can be obtained by introducing multi-level atoms and magnetic field gradients~\cite{douglas15a,gonzaleztudela15c,hung16a}. Apart from the photonic realizations~\cite{bravo12a,xie14a}, we foresee that these type of 2D structured reservoirs can be obtained in a wide variety of systems ranging from cold atoms~\cite{polini13a} in state dependent optical lattices~\cite{devega08a,navarretebenlloch11a} or 
circuit QED~\cite{houck12a}.

\acknowledgements{\emph{Acknowledgements.} The work of AGT and JIC was funded by the EU project SIQS and by the DFG within the Cluster of Excellence NIM. AGT also acknowledges support from Intra-European Marie-Curie Fellowship NanoQuIS (625955). We also acknowledge discussions with T. Shi, Y. Wu, S.-P. Yu, J. Mu\~{n}iz, and H.J. Kimble. }

\appendix

%
%
%
%
%
%
%
\setcounter{equation}{0}
\setcounter{figure}{0}
\makeatletter

\renewcommand{\thefigure}{A\arabic{figure}}
\renewcommand{\thesection}{\arabic{section}}  
\renewcommand{\theequation}{A\arabic{equation}}  

%
%

\section{Single QE dynamics  \label{secHC:single}}

As we show in the main text, the honeycomb like structure makes that finite size effects play a relevant role in the calculations. For that reason, we first show in Section~\ref{secHC:singleinf} the results associated to the continuum limit, $N\rightarrow\infty$, where one can obtain an analytical expression of $\Sigma_e(z)$. Afterwards, in Section~\ref{secHC:singlefin} we study how finite size effects correct the continuum predictions.

\subsection{Continuum limit \label{secHC:singleinf}}

The key point of this Section is that the analytical expression of $\Sigma_e(z)$ can be obtained by taking the continuum limit, that is,
\begin{align}
 \Sigma_e(z)=\frac{g^2}{N^2}\sum_\kk \frac{z}{z^2-|f(\kk)|^2}=\frac{g^2}{(2\pi)^2}\iint_{\mathrm{BZ}} d\kk \frac{z}{z^2-|f(\kk)|^2}\,.
\end{align}

\begin{figure*}
\centering
\includegraphics[width=0.98\linewidth]{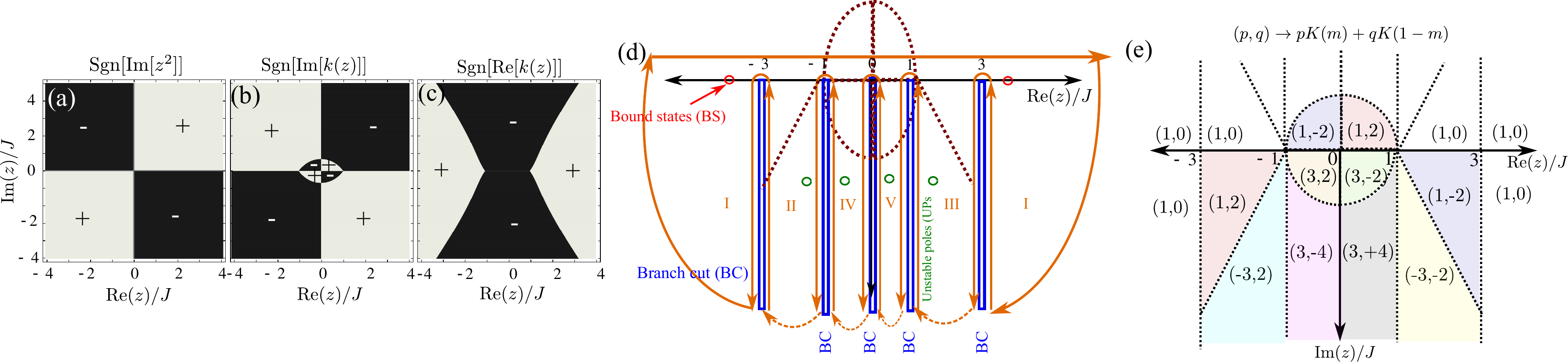}
\caption{(a-c) Sign of the $\mathrm{Im}[z^2]$, $\mathrm{Im}[k(z)]$ and $\mathrm{Re}(z)$ respectively that allows us to distinguish the different integrating regions required to perform the integration. (d) Contour of integration to calculate the probability amplitude $C_e(t)$ for the honeycomb bath. The self-energy is discontinuous at $\pm 3J,\pm J,0$ which force to take a detour and that give rise to branch cuts (BCs) contribution to the dynamics. We also denote in red/green the possible real/imaginary poles (BSs/UPs) appearing in the different Riemann sheets. (e) Combinations of coefficients $(p,q)$ of the elliptic integral $pK(m)+q K(1-m)$ which need to be used in the different coloured regions required for the integration contour of panel (d). }
\label{fig1HCL}
\end{figure*}

In this case, the $\kk$ integrals can be calculated exactly \cite{horiguchi72a}, first by demanding that $z\in \mathbb{R}$ and $z\notin [-3 J,3 J]$ and then, analytically continuing it to the rest of the complex plane. By doing so, one obtains the following expression:
\begin{align}
\label{eqHCL:self}
 \Sigma_e(z)&=\frac{g^2 z }{4\pi} C(z) \KK^{\mathrm{I}}(k(z)^2)\,, \\
 C(z)&=\frac{8}{(\sqrt{z^2}-J)^{3/2}(\sqrt{z^2}+3J)^{1/2}}\,,\\
 k(z)&=\frac{C(z) \sqrt[4]{z^2}}{2}
\end{align}
where $\KK^{\mathrm{I}}(m)$ has to be defined in a piecewise manner to guarantee analyticity in its definition domain:
\begin{align}
\KK^{\mathrm{I}}(m)&=\KK(m)\,,\,\mathrm{if}\,\,\mathrm{Im}[k(z)]\mathrm{Im}[z^2]<0\,,\nonumber\\
\KK^{\mathrm{I}}(m)&=\KK(m)+2i K(1-m)\,,\,\mathrm{if}\,\,\mathrm{Im}[k(z)],\mathrm{Im}[z^2]>0\,,\nonumber\\
  \KK^{\mathrm{I}}(m)&=\KK(m)-2i K(1-m)\,,\,\mathrm{if}\,\,\mathrm{Im}[k(z)],\mathrm{Im}[z^2]<0\,,
\end{align}
being $\KK(m)$ the complete elliptic integral of the first kind defined by:
\begin{equation}
\KK(m)=\int_0^{\pi/2}d\theta\frac{1}{\sqrt{1-m\sin^2(\theta)}}\,.
\end{equation}
The regions with different signs of $\mathrm{Im}[z^2]$ are $\mathrm{Im}[k(z)]$ plotted in Figs.\ref{fig1HCL}(a-b), together with the ones for the sign of $\mathrm{Re}[k(z)]$, which gives us the complete information about of the phase of the argument of the elliptic function [as $\mathrm{Im}[k^2(z)]=2\mathrm{Re}[k(z)]\mathrm{Im}[k(z)]$]. Evaluating $\Sigma_e(z)$ slightly above the real axis,  $\Sigma_e(E+i0^+)=\delta\omega_e(E)-i\frac{\Gamma_e(E)}{2}$, we obtain the functions $\delta\omega_e(E)$ and $\Gamma_e(E)$, that we plot in Fig.~\ref{fig1HC}(b). There, we observe that the function $\Gamma_e(E)$ [$\delta\omega_e(E)$] has a discontinuity [divergence] at $E=\pm 3J$ appearing at the border of the band edges of the upper/lower band, $\pm \omega(\kk)$. Moreover, at $E=\pm J$ the function also displays a divergence [discontinuity] at $E=\pm J$. Finally, at $E=0$ both functions are continuous, but have discontinuous derivative, therefore, being also non-analytical at that point.

From these functions we can already obtain the Markov prediction, that tells us that a single QE excitation decays at a rate $\Gamma_M=\Gamma_e(\Delta)$ to the bath. Thus, it predicts that: i) outside of the bands, i.e., $\Delta\notin [-3J,3J]$, or exactly at the Dirac point, $\Delta=0$, the QE should not decay because $\Gamma_M=0$; ii) Inside of the band, $\Delta\in [-3J,0)\cup (0,3J]$, the QE must decay with a rate $\Gamma_M$, which is infinite at $\Delta=\pm J$. However, we have seen in the main text, that these predictions get corrected when calculating the exact dynamics, so let us explain how these corrections appear within this formalism and which is their origin.

A typical way to obtain the exact dynamics of the probability amplitude $C_e(t)$ consists in calculating the Fourier Transform by closing the contour of integration in the lower plane of the complex plane ($\mathrm{Im}[z]<0$). However, when the $\Sigma_e(z)$ have regions where it is non-analytic, as the one of Eq.~\ref{eqHCL:self}, one must be careful when closing the contour to avoid them. One possible choice for the contour of integration is depicted in Fig.~\ref{fig1HCL}(d), where we take 5 detours of the integration contour at $E=\pm 3J, \pm J$ and $0$. Moreover, in the regions II-V, we analytically continue the function $\Sigma_e(z)$ into other Riemann sheets changing $\KK^\mathrm{I}(m)\rightarrow K^{\alpha}(m)$. The functions $\KK^\alpha(m)$ are obtained by imposing that $\Sigma_e^\alpha$ is continuum along the integration contour of Fig.~\ref{fig1HCL}. We exploit the fact that the elliptic functions are doubly periodic in which the different branches of $\KK(m)$ are given by combinations $p\KK(m)+qi K(1-m)$, with $p,q\in\mathbb{Z}$~\cite{abramowitz66a}. Using that, we obtain:
\begin{align}
 &\KK^{\mathrm{II}}(m)=(-3)K(m)+2iK(1-m)\,,\,\mathrm{if}\,, \mathrm{Re}[k(z)]>(<)0\,,\\
& \KK^{\mathrm{III}}(m)=(-3)K(m)-2iK(1-m)\,,\,\mathrm{if}\,, \mathrm{Re}[k(z)]>(<)0\,,\\
& \KK^{\mathrm{IV}}(m)=3K(m)+2(-4)iK(1-m)\,,\,\mathrm{if}\,, \mathrm{Im}[k(z)]>(<)0\,,\\
 &\KK^{\mathrm{V}}(m)=3K(m)(+4)-2iK(1-m)\,,\,\mathrm{if}\,, \mathrm{Im}[k(z)]<(>)0\,.\\
 \end{align}

 To make it more clear, in Fig.~\ref{fig1HCL}(e) we make a diagram with the different combinations of $(p,q)$ in all the regions relevant for the integration contour defined in Fig.~\ref{fig1HCL}(d). The results of the complete integration are shown in the main text.

\subsection{Finite system \label{secHC:singlefin}}

\begin{figure}
\centering
\includegraphics[width=0.85\linewidth]{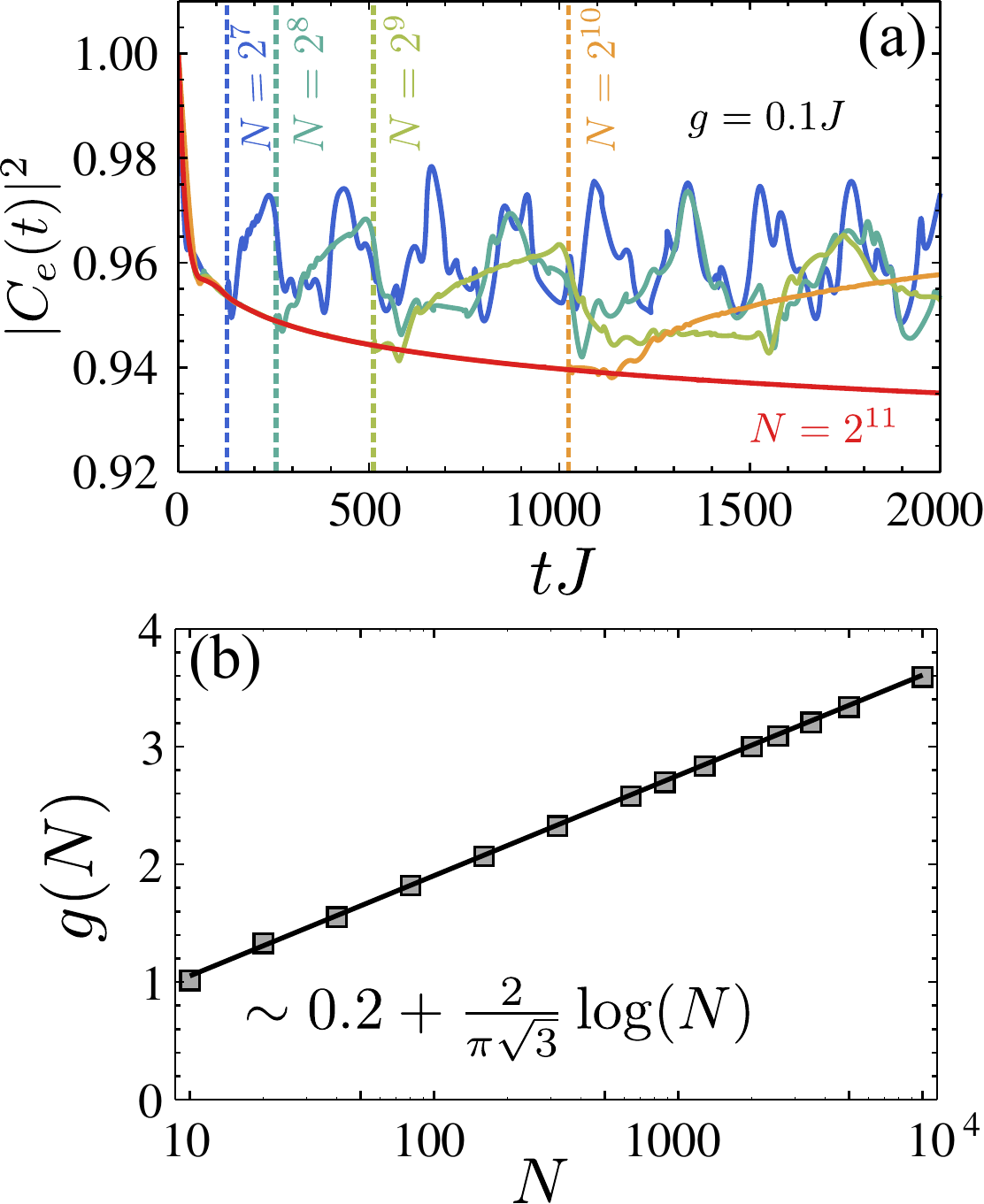}
\caption{(a) Dynamics of $|C_e(t)|^2$ for a single QE with $g=0.1J$ and different system sizes from $N=2^7$ to $N=2^{12}$ (see legend). (b) Numerical sum $g(N)$ defined in Eq.~\ref{eqHCL:gN} and its corresponding approximated expression.}
\label{fig3HCL}
\end{figure}

Due to the very slow decay of the excitations in the infinite case, it is expected that finite size effects play a relevant role in the discussion. In fact, we already show in Fig.~\ref{fig3HC} of the main text how the logarithmic spontaneous emission quenches after a finite time, $N$, and oscillates around a constant value which depends on the ratio $g/J$. In Fig.~\ref{fig3HCL}(a), we evidence that this effect depends strongly on the system size by plotting the decay of a QE with a fixed $g=0.1J$ for different number $N$ of bath modes. There, we observe that the quenching time appears at a longer timescale the larger the system size. To further characterize this phenomenon, we go back to the finite system self-energy expression, which reads:
\begin{equation}
 \Sigma_e(z)=\frac{g^2}{N^2}\sum_\kk \frac{z}{z^2-|f(\kk)|^2}\,.
\end{equation}

It is clear that $z=0$ is a solution from the pole equation, because $\Sigma_e(0)=0$. Its residue, which can be interpreted as the overlap of the initial wavefunction with the \emph{quasi}-bound state (qBS) emerging around the QE, is given by:
\begin{equation}
 R_0=\braket{\Psi_{\mathrm{qBS}}}{\Phi_0}=\frac{1}{1-\partial_z\Sigma_e(z)}\Big|_{z=0}=\frac{1}{1+\frac{g^2}{J^2} g(N)}\,,
\end{equation}
where $g(N)$ is a function that only depends on the system size $N$. This function can be approximated by going to the continuum limit and changing to the $\pp$-coordinates where the linear dispersion is isotropic in $\pp$ [$q_{1,2}=\frac{3}{2}\left(p_1\pm \frac{1}{\sqrt{3}} p_2\right)$], which leads:
\begin{equation}
 g(N)\approx\frac{2}{\sqrt{3}\pi}\int_{p_\mathrm{min}}^{p_c} dp\frac{1}{p}\,,
\end{equation}
where we include a minimum/maximum value for the region of integration to regularize the divergences of the integral. The minimum momentum is naturally provided by the discreteness of the lattice, i.e., $p_\mathrm{min}\propto \frac{1}{N}$. Thus, the $g(N)$ depends on the system size as follows:
\begin{equation}
 \label{eqHCL:gN}
 g(N)\approx C+\frac{2}{\pi\sqrt{3}}\log(N)\,.
\end{equation}
where $C$ is a constant that depends on the momentum cut-off $p_c$ and that we determine numerically by evaluating numerically the sum at a finite $N$, finding $C\approx 0.2$. Thus, the final residue of the qBS can be approximated by:
\begin{equation}
 R_0\approx \frac{1}{1+\frac{g^2}{J^2}(0.2+\frac{2}{\pi\sqrt{3}}\log(N))}\propto \frac{1}{\log(N)}\,,
\end{equation}
which goes to $0$ logarithmically with the systems size, $N$. The intuitive explanation for this is that for any finite system, the qBS is indeed square integrable as the rest of the eigenstates of the total Hamiltonian. Thus, the overlap with the initial wavefunction can be finite. However, for an infinite system the very slow decay of the qBS [$1/r$] makes it [marginally] not square integrable. Thus, it belongs to the continuous spectrum which has zero overlap with the initial wavefunction. In any practical situation, the relaxation dynamics will be a combination of a logarithmic decay for initial times, together with an absence of decay for long times due to the existence of the qBS. 

\section{Two QE dynamics \label{secHC:two}}

As we reviewed in the main text, the two QE dynamics simplifies because the symmetric and antisymmetric subspaces decouple. This is convenient because their corresponding probability amplitudes can be calculated through a similar Fourier transform than the one of a single QE but with a modified self-energy: $\Sigma^\beta_{\pm}(z;\nn_{12})=\Sigma_e(z)\pm \Sigma^\beta_{12}(z;n_{12})$, where the collective self-energy reads:
\begin{align}
\label{eqL:self}
\Sigma_{12}^\beta(z;\nn_{12})=\frac{g^2}{N^2}\sum_\kk \frac{D_\beta e^{i\kk\cdot\nn_{12}}}{z^2-|f(\kk)|^2}\,
\end{align}
where $D_\beta$ is a function which depends on whether the QE coupled to the $\beta=$AA/BB/AB sublattices. As we show in the main text $D_{\mathrm{AA}}=D_{\mathrm{BB}}=z$, whereas $D_{\mathrm{AB}}=D_{\mathrm{BA}}^*=f^*(\kk)$. In this case, it is difficult to obtain the analytical expression of the self-energy for every $z$ and $\nn_{12}$, and the expressions not be very enlightening~\cite{horiguchi72a}. Thus, we decide to follow the intuition developed from the single QE analysis to obtain some general expression for the $\Delta=0$ situation for both finite and infinite systems.

\subsection{Two QEs coupled to the AB sublattices}

The more attractive situation for coherent interactions is the case where the two QEs coupled to different sublattices, where:
\begin{align}
\label{eqL:selfAAB}
\Sigma_{12}^{AB}(z;\nn_{12})=\frac{g^2}{N^2}\sum_\kk \frac{f^*(\kk) e^{i \kk\cdot\nn_{12}}}{z^2-|f(\kk)|^2}\,.
\end{align}

To find an exact expression of $\Sigma_{12}^{AB}(z;\nn_{12})$ is hard because of the complicated shape of the energy dispersion $|f(\kk)|$. However, as in the previous Section one can obtain a closed simple expression for $\mathrm{Re}(z)/J\ll 1$ by taking it as the sum around the $\KKK_{\pm}$ points and making the transformation to the $\pp$-coordinates where the dispersion is isotropic. The contribution coming from the $\KKK_{\pm}$ point can be finally writen:
\begin{align}
\label{eqHCL:dev}
-i\frac{g^2}{J}\frac{e^{i\KKK_{\pm}\cdot\nn_{12}}}{(2\pi)^2} \frac{3\sqrt{3}}{2}\frac{2}{3}\hh_{p,\pm}^*\cdot\iint d\pp \frac{\pp}{\bar{z}^2-|\pp|^2}e^{i\pp \cdot \mm_{12}}\,,\end{align}
where $\hh_{p,\pm}=(-1,\pm i)$. Notice, we have redefined the $\bar{z}= 2 z/(3J)$ to make it adimensional and written a rescaled real space coordinates, $\mm_{12}=(m_1,m_2)=\left(\frac{3}{2}\left(n_1+n_2\right),\frac{\sqrt{3}}{2}\left(n_1-n_2\right)\right)$. From here, a final (assymptotic) expression can be obtained:
\begin{align}
\label{eqL:selfAAB}
\Sigma_{12}^{AB}(z;\nn_{12})&=i\frac{g^2}{J^2}\frac{z}{\sqrt{3}} H_1^1\left(\frac{2 z}{3J }|\mm_{12}|\right)\times\nonumber\\ &\times\frac{m_1\cos\left(\frac{2\pi}{3}\left(n_1-n_2\right)\right))-m_2\sin\left(\frac{2\pi}{3}\left(n_1-n_2\right)\right))}{|\mm_{12}|}\,,
\end{align}
 where $H_1^1(x)$ is the Hankel Function of the first kind. For small arguments, the Hankel function satisfies: $H_1(x)\approx -\frac{2 i}{\pi x}+O(x)$, such that:
 \begin{align}
\label{eqL:selfAABM}
\Sigma_{12}^{AB}(0;\nn_{12})&=J_M=\frac{g^2}{J}\frac{\sqrt{3}}{\pi |\mm_{12}|}\times\nonumber\\ &\times\frac{m_1\cos\left(\frac{2\pi}{3}\left(n_1-n_2\right)\right))-m_2\sin\left(\frac{2\pi}{3}\left(n_1-n_2\right)\right))}{|\mm_{12}|}\,.
\end{align}

For example, for $n_1=n_2=n$ as we are choosing for Fig.~\ref{fig4HC} of the main manuscript, where $\mm_{12}=(3n,0)$ and $|\mm_{12}|=3n$.
 \begin{align}
\label{eqL:selfAABMM}
\Sigma_{12}^{AB}(0;\nn_{12})&=J_M=\frac{g^2}{J}\frac{1}{\pi \sqrt{3} n}\propto \frac{1}{n}\,.
\end{align}

For $n_1=-n_2=n$, then $\mm_{12}=(0,\sqrt{3}n)$ and $|\mm_{12}|=\sqrt{3}n$, we have:
 \begin{align}
\label{eqL:selfAABMM}
\Sigma_{12}^{AB}(0;\nn_{12})&=J_M=-\frac{g^2}{J}\frac{1}{\pi  n}\sin\left(\frac{4\pi n}{3}\right)\,.
\end{align}
where $\sin\left(\frac{4\pi n}{3}\right)=-\frac{\sqrt{3}}{2},\frac{\sqrt{3}}{2},0,-\frac{\sqrt{3}}{2},\frac{\sqrt{3}}{2},0,\dots$.

Thus, within the Markov approximation the theory predicts coherent interactions decaying as a power law without any exponential attenuation, and with no imaginary component. However, we know from our previous experience with the single QE that one must be careful when extracting conclusions as non-Markovian corrections can be important around the Dirac point. Thus, in the next Section we deal with the exact problem and try to extract general conclusions.

\subsubsection{Exact integration}

We know that the pole equation for the symmetric/antisymmetric component for two QEs with $\Delta=0$ is given by:
\begin{equation}
\label{eqHCL:poleAB}
z_{\pm}=\Sigma^{AB}_{\pm}(z_{\pm};\nn_{12})=\frac{g^2}{N^2}\sum_\kk\frac{z_{\pm}\pm f^*(\kk)e^{i\kk \nn_{12}}}{z^2-|f(\kk)|^2}\,.
\end{equation}

In the main text, we show how for finite systems the solution to this pole equation is strictly real, that is, it is a real BS with no decay and which can give rise to oscillations [see Fig.~\ref{fig4HC}]. In order to further characterize it, we use the intuition developed from the previous Section which tell us that $\Sigma^{AB}_{\pm}(z;\nn_{12})$ goes to a function independent of $z$ as $z\rightarrow 0$. Thus, we can try to solve the previous Equation self-consistently by expanding the right-hand side of Eq.~\ref{eqHCL:poleAB} around $z=0$, arriving to:
\begin{equation}
z_{\pm}=\Sigma^{AB}_{\pm}(z_\pm;\nn_{12})\approx \Sigma_{12}^{AB}(0;\nn_{12})+ \partial_z\Sigma^{AB}_{\pm}(z;\nn_{12})\big|_{z=0} z_{\pm}+O(z^2)\,.
\end{equation}
which yields:
\begin{equation}
z_{\pm}\approx R_0 \Sigma_{12}^{AB}(0;\nn_{12})\,.
\end{equation}
where $R_0$ is the residue for the (quasi)-bound state that appears for a single QE at $\Delta=0$. Thus, the Markovian coherent interactions get renormalized by a factor $R_0\propto 1/\log(N)$, which goes to $0$ in the thermodynamic limit. However,  the convergence to zero with the system size is so slow that for any practical situation one can indeed observe the coherent interactions as shown in Fig.~\ref{fig4HC} of the main text. The last remaining thing is to calculate the associated residue of the symmetric/antisymmetric pole, which is given by:
\begin{equation}
R_{\pm}=\frac{1}{1-\partial_{z}\Sigma^{AB}_{\pm}(z)}\Big|_{z=z_{\pm}}\,.
\end{equation}

This is what we show in Fig.~\ref{fig4HC} of the main text, showing how it can still be large for large systems. To estimate the conditions under which this residue can be large, we can expand $\partial_{z}\Sigma^{AB}_{\pm}(z)$ for small $z$ [as we know $z_{\pm}\sim O\left(\frac{g^2}{J\log(N)}\right)$]:
\begin{align}
\label{eqHCL:exp2}
 \partial_{z}\Sigma^{AB}_{\pm}(z)\approx \partial_{z}\Sigma^{AB}_{\pm}(0)+\partial^2_{z}\Sigma^{AB}_{\pm}(z)|_{z=0} z\,.
\end{align}

It can be easily shown that $\partial_{z}\Sigma^{AB}_{\pm}(0)=-\frac{g^2}{J^2}g(N)$, that is the same function that controls the single QE residue. The only thing left to calculate is the contribution associated to the second derivative at $z=0$, which is given by:
\begin{align}
\partial^2_{z}\Sigma^{AB}_{\pm}(z)|_{z=0}=\mp 2\frac{g^2}{N^2}\sum_\kk \frac{f^*(\kk) e^{i\kk \nn_{12}}}{|f(\kk)|^4}\,.
\end{align}

Following similar recipes as the ones we have used for the rest of the results, that is, calculating the contribution around $\KKK_{\pm}$, changing to the $\pp$ coordinates where $f(\pp)$ is isotropic, we arrive to:
\begin{align}
\partial^2_{z}\Sigma^{AB}_{\pm}(z)|_{z=0}\propto \frac{g^2}{J^3}\int_{p_\mathrm{min}}^\infty dp \frac{J_1(p n)}{p^2}\sim -\frac{g^2}{J^3}\log(p_\mathrm{min})\,.
\end{align}

As the minimum momentum is set by: $p_\mathrm{min}=\frac{2\pi}{N}$ and $z_{\pm}\sim O\left(\frac{g^2}{J\log(N)}\right)$, then we find $\partial^2_{z}\Sigma^{AB}_{\pm}(z)|_{z=0} z_{\pm}\sim O\left(\frac{g^4}{J^4}\right)$. Thus, we can neglect the second term in the expansion of Eq.~\ref{eqHCL:exp2} as long as $g^2/J^2\ll 1$. Finally, the residue $R_{\pm}$ will be large as long as $g^2\log(N)/J^2\ll 1$.

\bibliography{Sci,Sciv2,books}

\end{document}